\documentclass[aps,prb,twocolumn,a4paper,showpacs,superscriptaddress,longbibliography]{revtex4-1}

\usepackage{amsmath,amssymb,bm,graphicx}
\usepackage{xcolor,changes}
\usepackage[colorlinks=true,urlcolor=blue,linkcolor=blue,citecolor=blue,bookmarks=false]{hyperref}

\begin{document}

\newcommand{\mfs}{Mn$_{1-x}$Fe$_{x}$Si}
\newcommand{\mcs}{Mn$_{1-x}$Co$_{x}$Si}
\newcommand{\mis}{Mn$_{1-x}$Ir$_{x}$Si}
\newcommand{\fcs}{Fe$_{1-x}$Co$_{x}$Si}
\newcommand{\cso}{Cu$_{2}$OSeO$_{3}$}


\title{Evolution of magneto-crystalline anisotropies in {\mfs} and {\mcs}\\as inferred from small-angle neutron scattering and bulk properties}

\author{J. Kindervater}
\affiliation{Physik-Department, Technische Universit\"{a}t M\"{u}nchen, D-85748 Garching, Germany}
 
\author{T. Adams}
 \affiliation{Physik-Department, Technische Universit\"{a}t M\"{u}nchen, D-85748 Garching, Germany}
 
\author{A. Bauer}
 \affiliation{Physik-Department, Technische Universit\"{a}t M\"{u}nchen, D-85748 Garching, Germany}
 
\author{F. Haslbeck}
 \affiliation{Physik-Department, Technische Universit\"{a}t M\"{u}nchen, D-85748 Garching, Germany}

\author{A. Chacon}
 \affiliation{Physik-Department, Technische Universit\"{a}t M\"{u}nchen, D-85748 Garching, Germany}

\author{S. M\"{u}hlbauer}
 \affiliation{Heinz Maier-Leibnitz Zentrum (MLZ), Technische Universit\"{a}t M\"{u}nchen, D-85748 Garching, Germany}

\author{F. Jonietz}
 \affiliation{Physik-Department, Technische Universit\"{a}t M\"{u}nchen, D-85748 Garching, Germany}

\author{A. Neubauer}
 \affiliation{Physik-Department, Technische Universit\"{a}t M\"{u}nchen, D-85748 Garching, Germany}

\author{U. Gasser}
 \affiliation{Laboratory for Neutron Scattering and Imaging, Paul Scherrer Institut, CH-5232 Villigen, Switzerland}

\author{G. Nagy}
 \affiliation{Laboratory for Neutron Scattering and Imaging, Paul Scherrer Institut, CH-5232 Villigen, Switzerland}

\author{N. Martin}
 \altaffiliation[Present address: ]{Laboratoire L{\'e}on Brillouin, CEA, CNRS, Universit{\'e} Paris-Saclay, CEA Saclay 91191 Gif-sur-Yvette, France }
 \affiliation{Physik-Department, Technische Universit\"{a}t M\"{u}nchen, D-85748 Garching, Germany}
 \affiliation{Heinz Maier-Leibnitz Zentrum (MLZ), Technische Universit\"{a}t M\"{u}nchen, D-85748 Garching, Germany}
  
\author{W. H\"{a}u{\ss}ler}
 \affiliation{Physik-Department, Technische Universit\"{a}t M\"{u}nchen, D-85748 Garching, Germany}
 \affiliation{Heinz Maier-Leibnitz Zentrum (MLZ), Technische Universit\"{a}t M\"{u}nchen, D-85748 Garching, Germany}
 
\author{R. Georgii}
 \affiliation{Physik-Department, Technische Universit\"{a}t M\"{u}nchen, D-85748 Garching, Germany}
 \affiliation{Heinz Maier-Leibnitz Zentrum (MLZ), Technische Universit\"{a}t M\"{u}nchen, D-85748 Garching, Germany}
 
\author{M. Garst}
 \affiliation{Institute for Theoretical Physics, Universit\"{a}t zu K\"{o}ln, D-50937 K\"{o}ln, Germany}
 \affiliation{Institut f\"{u}r Theoretische Physik, Technische Universit\"{a}t Dresden, D-01062 Dresden, Germany}

\author{C. Pfleiderer}
 \affiliation{Physik-Department, Technische Universit\"{a}t M\"{u}nchen, D-85748 Garching, Germany}
 
\date{\today}

\begin{abstract}
We report a comprehensive small-angle neutron scattering~(SANS) study of magnetic correlations in Mn$_{1-x}$Fe$_{x}$Si at zero magnetic field. To delineate changes of magneto-crystalline anisotropies (MCAs) from effects due to defects and disorder, we recorded complementary susceptibility and high-resolution specific heat data, and investigated selected compositions of {\mcs}. 
For all systems studied the helimagnetic transition temperature and magnetic phase diagrams evolve monotonically with composition consistent with literature. 
The SANS intensity patterns of the spontaneous magnetic order recorded under zero-field cooling, which were systematically tracked over forty angular positions, display strong changes of the directions of the intensity maxima and smeared out intensity distributions as a function of composition.
We show that cubic MCAs account for the complex evolution of the SANS patterns, where for increasing $x$ the character of the MCAs shifts from terms that are fourth-order to terms that are sixth order in spin--orbit coupling.
The magnetic field dependence of the susceptibility and SANS establishes that the helix reorientation as a function of magnetic field for Fe- or Co-doped MnSi is dominated by pinning due to defects and disorder. 
The presence of well-defined thermodynamic anomalies of the specific heat at the phase boundaries of the skyrmion lattice phase in the doped samples and properties observed in {\mcs} establishes that the pinning due to defects and disorder remains, however, weak and comparable to the field scale of the helix reorientation.
The observation that MCAs, that are sixth order in spin-orbit coupling, play an important role for the spontaneous order in {\mfs} and {\mcs}, offering a fresh perspective for a wide range of topics in cubic chiral magnets such as the generic magnetic phase diagram, the morphology of topological spin textures, the paramagnetic-to-helical transition, and quantum phase transitions.
\end{abstract}


\maketitle


\section{Introduction}
\subsection{Motivation}

Cubic chiral magnets such as MnSi, FeGe, and {\fcs} have been attracting great scientific interest for many decades. Seminal work in the early 1980s established these materials as the first examples stabilizing incommensurate modulated magnetic order consistent with so-called Lifshitz invariants~\cite{dzialoshinskii:57, moriya:60, dzialoshinskii:64, cummins:90, 1976:Ishikawa:SolidStateCommun}. At the same time, pioneering studies identified members of the same material class as prototypical representatives of weak itinerant-electron magnetism, where the subsequent development of a self-consistent theory taking into account the effects of spin fluctuations was found to be in excellent agreement with experiment~\cite{1985:Lonzarich:JPCSP, 1985:Moriya:book}. This set the stage for the first high-pressure studies of quantum phase transitions (QPTs) and the discovery of putative evidence suggesting a generic breakdown of Fermi liquid theory~\cite{1993:Pfleiderer, 1997:Pfleiderer:PhysRevB, 2007:HvL:RMP}. On a rather different note, skyrmions as the first example of topological spin textures could be identified in recent years in the same materials, opening a new avenue for spintronics applications~\cite{2009:Muhlbauer:Science, 2010:Yu:Nature, 2011:Yu:NatureMater, 2012:Seki:Science, 2013:Nagaosa:NN}. Despite their long history, the properties of cubic chiral magnets represent an active topic of present-day research comprising several independent directions. For many of these areas the character and strength of magneto-crystalline anisotropies (MCAs) are of great interest. 

An important aspect of the different areas of research pursued in the class of cubic chiral magnets is the presence of a well-developed hierarchy of energy scales~\cite{1980:Landau:Book}. On the strongest scale, exchange interactions favor parallel spin alignment. This is followed, on intermediate scales, by Dzyaloshinsky--Moriya spin--orbit interactions (DMI) permitted in the non-centrosymmetric crystal structure (space group $P2_ {1}3$). The weakest energy scale are, finally, higher-order spin--orbit coupling terms, also known as cubic MCAs. To leading order these comprise contributions that are fourth and sixth order in spin--orbit coupling~(SOC). As the fourth-order terms in SOC by default are expected to be stronger, present-day investigations have almost exclusively considered this part only. In contrast, the importance of the sixth-order terms in SOC has only recently been addressed in a few selected studies as summarized further below.

Numerous studies have shown, that substitutional doping of MnSi with Fe or Co sensitively suppresses the onset of long range magnetic order. In turn, {\mfs} and {\mcs} allow to put the validity of the description of MnSi and the class of cubic chiral magnets to a critical test as long-range helical order collapses, shedding light on the paramagnetic to helimagnetic transition, the generic appearance of the magnetic phase diagram, the evolution of the skyrmion lattice \cite{2014:Franz:PhysRevLett}, and the properties of QPTs \cite{2010:Bauer:PhysRevB}.  This raises the question for the evolution of the MCAs in {\mfs} and {\mcs} as a key property controlling the nature of the microscopic magnetic structure as well as the formation of domains. However, as doping generates also defects and disorder, an important question concerns the influence of defect-related pinning on the long-range character of the magnetic properties and the associated phase transitions.

In this paper we report an investigation of the microscopic evolution of the magnetic order in {\mfs} up to $x_{\rm Fe} = 0.10$, complemented by further data in {\mfs} up to $x_{\rm Fe} = 0.22$ and selected compositions $x_{\rm Co}$ in {\mcs}. For our study we combined comprehensive small-angle neutron scattering experiments with measurements of the magnetization, ac susceptibility, and specific heat. Our methodological approach focusses on the determination of the evolution of spontaneous magnetic order under zero-field cooling as a function of substitutional doping in {\mfs} and {\mcs}. In addition, we performed measurements of the magnetic field dependence of the magnetic order to gain insights on the interplay of the Zeeman energy with the MCAs and the effects of pinning and disorder.

As our main result we find a complex variation of the SANS intensity patterns of the magnetic order in {\mfs} and {\mcs} under zero-field cooling as a function of composition. We find further that the effects of defects and disorder in combination with the MCAs control the helix reorientation under magnetic field. We show that conventional cubic MCAs account for all of our experimental findings, where for increasing $x$ the character of the MCAs shifts from terms that are fourth order to terms that are sixth order in spin-orbit coupling. This connects with numerous scientific questions pursued in the class of cubic chiral magnets.


\subsection{Terminology and Outline}

For the sake of accessibility we present in the following a detailed account of the chain of arguments and the outline of our paper. It proves to be helpful to begin this account with a brief introduction to the current understanding of the magnetic order in defect-free cubic chiral magnets and the terminology used to describe these properties.

It has long been established that the hierarchical energy scales in cubic chiral magnets, in the most general case, lead to the formation of long-wavelength helicoidal modulations~\cite{1980:Bak:JPhysCSolidState, Nakanishi:SSC1980, 2016:Bauer:Book}.  For increasing order of length scales the magnetic properties of bulk samples comprise (i) the size and the orientation of the magnetization on atomic scales, (ii) the wavelength, $\lambda$, and harmonicity of the helicoidal twisting, (iii) the direction of the helicoidal modulation vector $\vec{Q}$ in the presence and population of magnetic domains.

Under zero field cooling and below a transition temperature $T_c$ the helicoidal modulations in MnSi form long-range helical order. That is, there are equal domain populations for propagation directions $\vec{Q}$ along the $\langle 111\rangle$ easy axes, the twisting exhibits a simple sinusoidal helical modulation, the magnitude of the moments is constant, and the moments are oriented perpendicular to $\vec{Q}$.

Application of a magnetic field affects the magnetic properties in terms of the Zeeman energy, i.e., the field changes the  magnitude and orientation of the magnetization, the direction and harmonicity of the helicoidal modulation, and the domain populations. Above a critical magnetic field $H_{c1}$, the modulation vector $\vec{Q}$ aligns parallel to the applied field, and the magnetic moments feature an angle less than $90^{\circ}$ with respect to the direction of $\vec{Q}$. In other words, an applied magnetic field stabilizes a uniform component of the magnetization parallel to the field direction. Above the so-called upper critical field, $H_{c2}$, the modulation collapses, resulting in a field-polarized (ferromagnetic) state.

In case of a zero-field-cooled state the magnetic state for magnetic fields below $H_{c1}$ corresponds to a multi-domain helicoidal configuration. For the sake of convenience this state is generally referred to as helical order in the literature. Further, for magnetic fields in the range $H_{c1}< H <H_{c2}$ the modulations are referred to as conical state. Thus, the characteristic field at $H_{c1}$ is widely known as the helical to conical transition, a terminology ignoring the details of the magnetic order below $H_{c1}$ as well as the character of the reorientation.

Following seminal work in MnSi by Grigoriev \cite{2006:Grigoriev:PhysRevB2} and Maleyev \cite{2006:Maleyev:PRB}, on the helix reorientation, recent studies of the magnetization suggest the formation of a single domain state under field cycling across $H_{c1}$ for field along the $\langle 111\rangle$ \cite{2017:Bauer:PRB,2013:Narozhnyi:JETP,2015:Narozhnyi:PRB}. An analogous single domain state was recently observed in {\cso} under field cycling along $\langle 100\rangle$ \cite{2018:Chacon:NatPhys,2018:Halder:PRB}. A comprehensive account was finally established in recent high-precision small-angle neutron scattering, magnetization, and ac susceptibility measurements of the helicoidal modulations and associated evolution of domain populations across $H_{c1}$, which are fully accounted for in terms of leading order MCAs that are fourth and sixth order in SOC \cite{2017:Bauer:PRB}. The main effect of a magnetic field on the helicoidal order is twofold: (i) it reorients the helix and thus the vector $\vec{Q}$, and, (ii) it induces a finite magnetic polarization, which, in general, is not aligned with the applied magnetic field as reflected by the asymmetry of the susceptibility tensor. The modulus of $\vec{Q}$, the magnitude of the magnetic moment and the harmonicity of the magnetic modulation are essentially not affected. Moreover, depending on crystallographic orientation the value of $H_{c1}$ varies substantially and the characteristics at $H_{c1}$ changes, displaying a cross-over, or a first or a second order transition.

Just below the helical transition temperature $T_c$ entropic effects associated with thermal fluctuations stabilize the skyrmion lattice phase in a small magnetic field range~\cite{2009:Muhlbauer:Science, 2010:Munzer:PhysRevB, 2013:Buhrandt:PRB, 2013:Nagaosa:NN}. To leading order the skyrmion lattice may be approximated by three helical modulations perpendicular to the applied magnetic field, where the propagation directions enclose angles of $120^{\circ}$~\cite{2009:Muhlbauer:Science, 2011:Adams:PhysRevLett}. In the bulk properties of MnSi the effects of MCAs result in changes of the temperature range of the skyrmion lattice phase by a factor of two, extending over roughly 1\,K for the $\langle111\rangle$ easy axes and roughly 2\,K for the $\langle100\rangle$ hard axes. When the field is not aligned along a crystallographic high-symmetry direction, the MCAs induce also a slight tilting of the skyrmion lattice plane with respect to the field direction as well as the in-plane orientation of the skyrmion lattice with respect to the crystal lattice~\cite{2009:Muhlbauer:Science, 2010:Munzer:PhysRevB, 2010:Adams:JPCS, 2018:Adams:PRL, Adams:PhD-thesis, 2017:Bannenberg:PRB}.

In this paper we focus on the character of spontaneous magnetic correlations in {\mfs} and {\mcs}, i.e., at zero magnetic field, using small angle neutron scattering. As our main observation we find an unexpectedly complex evolution of the intensity pattern as a function of composition. As previous studies had shown a monotonic evolution of the helimagnetic transition temperature and the phase boundaries of the magnetic phase diagram under doping, an important aspect concerns changes of the MCAs and the importance of pinning due to defects and disorder.

It proofs to be helpful to begin the presentation of our results with an account of the bulk properties in Sec.~\ref{ResultsBulk}. On the one hand, this allows to emphasize systematic variations not reported before, while demonstrating consistency with previous studies. On the other hand, the bulk properties set the stage for the presentation of the microscopic insights obtained in SANS for selected temperatures and magnetic fields as the main focus of our study.

The presentation of the bulk properties focuses on the signatures of the helix reorientation and the temperature range of the skyrmion lattice phase, representing the two aspects where the effects of MCAs are most pronounced. Namely, in pure MnSi the helical to conical transition at $H_{c1}$ is connected with distinct signatures both under zero-field cooling and field cooling, featuring the MCA for field along $\langle 100\rangle$, $\langle 110\rangle$, and $\langle 111\rangle$. Deviations between $\mathrm{d}M/\mathrm{d}H$ and $\mathrm{Re}\chi_{\mathrm AC}$ additionally reflect the characteristics of slow dynamical processes\cite{2012:Bauer:PhysRevB,2016:Bauer:PhysRevB}.

As reported in Sec.~\ref{mfs-susc}, under Fe doping exceeding $x_{\mathrm{Fe}}\geq0.04$ the signatures of $H_{c1}$ in $\mathrm{d}M/\mathrm{d}H$ and $\mathrm{Re}\chi_{\mathrm AC}$ are only observed after initial zero-field cooling, whereas no signatures are observed after field cycling regardless of crystallographic orientation. Moreover, for $x_{\mathrm{Fe}}\geq0.04$ values of $H_{c1}$ observed under ZFC are the same for different crystallographic orientations. With increasing $x_{\mathrm{Fe}}$ the values of $H_{c1}$ as extrapolated to zero temperature, increase by a factor of two and $H_{c1}$ decreases with increasing temperature as opposed to the lack of a significant temperature dependence for $x_{\mathrm{Fe}} <0.04$. 
Similar changes are also seen for the skyrmion lattice phase, where we find that the dependence on crystallographic orientation of the temperature range vanishes for $x_{\mathrm{Fe}} \geq0.04$. Taken together this suggests for $x_{\mathrm{Fe}}\geq0.04$, that isotropic pinning due to defects and disorder dominates as compared to the MCAs alone.

Several studies reported spin glass behavior in polycrystalline samples of {\mfs} and {\mcs} for large doping levels\cite{2010:Teyssier:PhysRevB,Demishev:2016}. In turn, the behavior at $H_{c1}$ raises the question how to better gauge the effects of defects and disorder as compared with the hierarchy of energy scales, and if there are still well-defined thermodynamic phase boundaries and long-range order under increasing doping. To clarify this question we pursued two different approaches. As reported in Sec.~\ref{mfs-specheat}, we performed high-resolution specific heat measurements of the skyrmion lattice phase under Fe-doping, representing a fragile state confined to a small parameter regime. Here we find well-defined anomalies of the transition to the skyrmion lattice order, suggesting that the thermodynamic phases remain well-defined under doping for $x<x_c$.

Moreover, we considered the evidence for defect-related pinning of selected compositions of {\mcs}, reported in Sec.~\ref{mcs-susc}. Previous studies had shown that the same changes of the transition temperature and ordered moment under Fe-doping may be observed for half of the Co concentration in {\mcs}, consistent with the difference of valence electron count of Mn under Fe- and Co-doping\cite{2010:Bauer:PhysRevB}. While the amount of disorder under Co-doping is a factor of two smaller as compared with Fe-doping, we find that the effects of defect-related pinning set in at a lower Co concentration. This implies that the relative importance of defects and disorder with respect to the MCAs becomes stronger, while the total strength of the MCAs must be moderately decressing under increasing doping. Thus both, the specific heat measurements in {\mfs} and the studies of {\mcs}, demonstrate that the effects of pinning and disorder are comparable to the MCAs such that the associated thermodynamic phase transitions and long-range order are still well-defined. 

As the behavior at $H_{c1}$ under Fe-doping is dominated by pinning due to defects and disorder, the helix reorientation may no longer be used to infer information on the nature and strength of the MCAs as for pure MnSi. Instead microscopic information on the domain distribution and helicoidal order are required. Reported in Sec.~\ref{ResultsSANS} are the results of detailed SANS studies of {\mfs} and {\mcs}. Combining data recorded under forty different angular orientations, we report in Sec.~\ref{mfs-SANS} the observation of strong changes of the direction of the modulation $\vec{Q}$ of the zero-field cooled ground state under increasing Fe-doping, whereas the modulus of $\vert\vec{Q}\vert$ decreases monotonically with increasing $x_{\rm Fe}$. Analogous changes are observed under Co-doping as reported in Sec.~\ref{mcs-SANS}, where changes of $\vert\vec{Q}\vert$ as a function of composition are consistent when $x_{\mathrm{Co}}$ is about half of $x_{\mathrm{Fe}}$.

Further, following application of a sufficiently large magnetic field, a single-domain conical state survives for decreasing field below $H_{c1}$ as summarized in Sec.~\ref{Hc1-SANS}. Finally, as reported in Sec.~\ref{skx-SANS}, a well-defined diffraction pattern characteristic of long-range order is observed in the skyrmion lattice phase as anticipated from the specific heat anomalies. Thus, the combination of bulk properties and SANS data consistently establish that the effects of defects and disorder associated with substitutional doping in {\mfs} and {\mcs} lead to pinning comparable in strength with the MCAs. In comparison, the modulation length $\vert\vec{Q}\vert$, which reflects the competition of ferromagnetic exchange and DMI, displays a moderate monotonic variation as a function of composition and does not seem to be sensitive to disorder.

Our experimental data raise the question for the origin of the changes of propagation direction of the helicoidal modulations at zero field and the implications of these findings. The discussion reported in Sec.~\ref{discussion} addresses these aspects. Starting with the same MCAs considered in pure MnSi as an account of the helix reorientation and the detailed orientation of the skyrmion lattice order, we show in Sec.~\ref{disc-MAC} that the changes of easy axis in {\mfs} and {\mcs} may be explained in terms of a change of the character of the MCAs from terms that are fourth order in SOC to terms that are sixth order in SOC. This is followed in Sec.~\ref{disc-broad} by a discussion of the implications of this observation for the broader understanding of cubic chiral magnets. Our paper closes with a set of conclusions presented in Sec.~\ref{Conclusion}.


\section{Broader Context}
\label{extintro}

The extended introduction presented in this section serves to provide more detailed information on the different areas of research for which the insights gained in our studies of {\mfs} and {\mcs} are of interest. The entire section was born out of the refereeing process of this paper, which highlighted the lack of a suitable review paper on the role and importance of MCAs in cubic chiral magnets. It may be readily skipped when familiar with this field of research. The section begins with an account of further  aspects of the magnetic phase diagram of cubic chiral magnets in Sec.~\ref{genericpd}, followed by the paramagnetic-to-helimagnetic transition in Sec.~\ref{paraheli}, the properties of the QPT in MnSi in Sec.~\ref{QPT}, and previous studies of {\mfs} and {\mcs} in Sec.~\ref{mfsmcs}. The section closes in Sec.~\ref{microsc} with a short account of studies addressing the microscopic origin of the different interactions. 


\subsection{Generic magnetic phase diagram}
\label{genericpd}

The hierarchical set of energy scales in the class of cubic chiral magnets is directly reflected in the generic magnetic phase diagram as summarized above~\cite{1980:Bak:JPhysCSolidState, Nakanishi:SSC1980, 2016:Bauer:Book}. Well-established are the formation of multi-domain helical order at zero magnetic field, the helix reorientation at $H_{c1}$, the suppression of the conical modulation above $H_{c2}$, and the skyrmion lattice phase in the vicinity of the paramagnetic-to-helimagnetic transition. As a function of crystallographic orientation, the temperature and field range of this high-temperature skyrmion lattice phase is nearly isotropic.

In contrast, early theoretical calculations suggested, that skyrmion lattice order could be stabilized by magneto-crystalline uniaxial anisotropies which are second order in SOC and not permitted in cubic chiral magnets~\cite{1989:Bogdanov, 1994:Bogdanov}. By now, several materials with reduced crystal symmetry have been identified, which feature skyrmion lattices stabilized by these uniaxial magneto-crystalline anisotropies~\cite{2015:Kezsmarki:NM, 2017:Nayak:Nature}. Recent studies in {\cso} unexpectedly revealed that certain fourth-order contributions in SOC to the MCAs may become dominant close to $H_{c2}$, where they drive the formation of two new phases for the $\langle100\rangle$ direction only, notably a tilted conical state and a low-temperature skyrmion phase~\cite{2018:Chacon:NatPhys,2018:Halder:PRB,2018:Qian:SciAdv}. Thus, the precise orientation of both the conventional high-temperature skyrmion lattice and the low-temperature skyrmion phase offer important consistency checks on the MCAs inferred from the spontaneous magnetic order and the helix reorientation. 


\subsection{Paramagnetic-to-helimagnetic transition}
\label{paraheli}

The weak interaction scales associated with higher-order SOC play also a decisive role for the character of the fluctuations in the paramagnetic state of the cubic chiral magnets and for the precise nature of the helimagnetic phase transition~\cite{2013:Janoschek:PhysRevB}. Well above the helimagnetic transition temperature, deep in the paramagnetic state, these fluctuations correspond to conventional (ferromagnetic) paramagnons dominated by the strongest scale, i.e., the ferromagnetic exchange interactions. With decreasing temperature the relative importance of the DMI, representing the intermediate scale, gain weight and the fluctuations are enhanced on the length scale of the helical modulation. In turn, the fluctuations dominate on the surface of a small sphere in reciprocal space, i.e., the volume in reciprocal space accessible for magnetic correlations increases resulting in a strong enhancement of the fluctuations.  

As long as the cubic MCAs are weak, the concomitant enhancement of the fluctuations suppresses the helimagnetic order until a fluctuation-induced first-order transition takes place at $T_{c}$, leaving behind a fluctuation-disordered (FD) regime below a cross-over, $T_{2}$. This so-called Brazovskii scenario is in excellent agreement with neutron scattering, specific heat and ultrasound attenuation~\cite{1975:Brazovskii:SovPhysJETP, 2007:Stishov:PhysRevB, 2008:Stishov:JPhysCondensMatter, 2009:Petrova:JPhysCondensMatter, 2013:Janoschek:PhysRevB, Pappas:PRL2009, 2014:Kindervater:PhysRevB, 2014:Zivkovic:PRB,Nii:2014}. 

In contrast, when the cubic MCAs are strong, the presence of the fluctuations may be reduced to specific directions before the transition takes place. Depending on the remaining phase space available, a fluctuation-induced first-order transition may still be expected according to a well-known scenario proposed by Bak and Jensen~\cite{1980:Bak:JPhysCSolidState}. Under an applied magnetic field the fluctuations will eventually get quenched sufficiently such that the transition changes from first to second order at a well-documented field-induced tricritical point~\cite{2013:Bauer:PhysRevLett, 2016:Stishov:PRB}. Recent neutron scattering studies  have been portrayed to question the Brazovskii scenario and associated tricritical point \cite{Pappas:PRL2017}. However, this paper does not offer any insights on the strength of the MCAs at the heart of the Brazovskii-scenario, and does not offer an alternative explanation either. Thus, further evidence on the strength and character of the MCAs represent a key aspect of understanding the thermodynamic nature of the paramagnetic-to-helimagnetic transition.  


\subsection{Quantum phase transitions under pressure}
\label{QPT}

A major field of research which epitomizes the importance of fluctuations and stochastic processes for the formation of novel forms of order concerns QPTs, i.e., phase transitions at zero temperature. Motivated by the sensitivity to pressure, selected cubic chiral magnets have been among the first materials in which QPTs were studied under controlled conditions using hydrostatic pressure~\cite{1993:Pfleiderer, 1997:Pfleiderer:PhysRevB, 2012:Petrova:PRB}. Based on the excellent account of their magnetic properties at ambient pressure as compared with other strongly correlated materials, these studies have become an important point of reference for the general understanding of QPTs in $d$- and $f$-electron materials~\cite{2007:HvL:RMP,2015:Schmakat:EPJSP}.

A prominent unresolved question represents the nature of a $T^{3/2}$ non-Fermi liquid~(NFL) temperature dependence of the electrical resistivity in MnSi and FeGe above a critical pressure $p_{c}$ at which $T_{c}$ is suppressed~\cite{1997:Pfleiderer:PhysRevB, 2001:Pfleiderer:Nature, 2003:Doiron-Leyraud:Nature, 2006:Pedrazzini:PhysicaB, 2006:Petrova:PhysRevB2, 2007:Pfleiderer:Science,2015:Barla:PRL}. Detailed studies of the Hall effect connect the regime of the NFL resistivity with topological spin textures, namely skyrmions. However, as muon spin rotation ($\mu$-SR) and nuclear magnetic resonance (NMR) studies in MnSi did not detect static magnetic moments these textures must be dynamic~\cite{2013:Ritz:PhysRevB, 2013:Ritz:Nature, 2007:Uemura:NaturePhys, 2004:Yu:PhysRevLett}. Yet, neutron scattering, at least in part of the temperature versus pressure range of the NFL resistivity, revealed substantial scattering intensity on the surface of a sphere in reciprocal space with a radius corresponding to the helical pitch~\cite{2004:Pfleiderer:Nature,2005:Fak:JPhysCondMatter}. Alluding to intensity patterns detected in liquid crystals, this observation has been interpreted as partial magnetic order, where the time scales probed in neutron scattering, as compared with $\mu$-SR and NMR, imply a dynamic state in the range between $10^{-10}$~s and $10^{-11}$~s. 

Keeping in mind the role of MCAs for the magnetic order at ambient pressure, where inherently different phases form in terms of the character and directions of the modulations with wavevector $Q$, the nature of the partial magnetic order has created great interest. A seminal proposal concerned, for instance, an abundance of magnetic quantum rotons in the spirit of the Brazovskii transition that is also relevant for the paramagnetic-to-helimagnetic transition at ambient pressure~\cite{2004:Schmalian:PRL}. However, in the partially ordered regime most unusual is the observation of broad neutron scattering maxima along the crystalline $\langle110\rangle$ directions, which could not be reconciled with the fourth-order SOC contributions considered so far only. This observation motivated the proposal of complex multi-$Q$ textures, referred to as spin crystals, which provided the theoretical framework used in the identification of the skyrmion lattice order at ambient pressure~\cite{2006:Binz:PRL, 2006:Binz:PRB}.

The observation of partial order at the QPT raises the question for the precise nature of the MCAs under pressure. Indeed, basic susceptibility measurements and neutron scattering of the magnetic field dependence suggested unchanged values of $H_{c1}$ and $H_{c2}$ of the helimagnetic order under pressure, broadly implying unchanged MCAs~\cite{1997:Thessieu:JPhysCondensMatter, 2007:Pfleiderer:PhysRevLett,2019:Bannenberg:PhysRevB}. However, recent progress made on the magnetic phase diagram at ambient pressure as well as the results reported in this paper strongly motivate to revisit the MCAs at the pressure-tuned QPT of MnSi and FeGe.


\subsection{Properties of {\mfs} and {\mcs}}
\label{mfsmcs}

The helimagnetic transition temperature in cubic chiral magnets responds sensitively to substitutional doping~\cite{1983:Beille:SolidStateCommun, 1998:Achu:JMagnMagnMater}. In turn, compositional changes represent an important tool when trying to identify the microscopic origin of the hierarchy of energy scales and the effects of fluctuations in cubic chiral magnets. A question that has not been addressed consequently concerns, however, the presence of disorder and defects as well as the putative presence of compositional gradients and inhomogeneities. From an experimental point of view these require additional caution.

Early studies of the electrical resistivity and M\"{o}ssbauer spectroscopy in polycrystalline {\mfs} provided first evidence of the suppression of helimagnetic order under doping~\cite{1983:Waki:JMMM, 1984:Nishihara:PRB}. Likewise, early magnetization measurements and NMR in polycrystalline {\mcs} identified a suppression of the helimagnetic order with increasing cobalt content~\cite{1978:Motoya:JPhysSocJpn}. Subsequent small-angle neutron scattering in polycrystalline {\mfs} and {\mcs} revealed a moderate reduction of the helical modulation length with increasing $x$ without comparison between iron and cobalt doping~\cite{1983:Beille:SolidStateCommun, 1998:Achu:JMagnMagnMater}. 

The evolution of the MCAs in {\mfs}, as inferred from the magnetic phase diagrams determined by small-angle neutron scattering and selected magnetization measurements, was addressed for $x_{\rm Fe} = 0.06$, 0.08, and 0.10 in Ref.~\onlinecite{2009:Grigoriev:PhysRevB}. Based on SANS data for $x_{\rm Fe} = 0.08$, the authors suggested qualitatively that the orientation of the magnetic helix axis changes from $\langle111\rangle$ to $\langle100\rangle$ with streaks of intensity connecting the two directions. However, sixth-order contributions in SOC were not considered. Further, the ratio of the DMI to ferromagnetic exchange was inferred from the helical modulation length, whereas the strength of the magnetic anisotropies was taken from the field value, $H_{c1}$, of the helix reorientation, assuming that $H_{c1}$ is not affected by disorder. As their main conclusion, the authors suggested that the ferromagnetic exchange softens for a critical iron composition of $0.12$, representing the main mechanism controlling the suppression of the helimagnetic transition. However, such a softening would be inconsistent with the small change of the fluctuating moment of the Curie-Weiss dependence observed under doping \cite{2010:Bauer:PhysRevB}. In the study reported in the present paper we go experimentally well beyond the work reported in Ref.~\onlinecite{2009:Grigoriev:PhysRevB}, arriving at different conclusions.

The putative existence of QPTs in {\mfs} and {\mcs} was pursued in several studies starting with measurements of the resistivity~\cite{2009:Meingast:Book}. Magnetization, susceptibility, and specific heat data consistently suggested that the magnetic phase diagram remains essentially unchanged, while the helimagnetic transition temperature $T_{c}$ vanishes above a critical composition ($x_{\mathrm{Fe}} > 0.10$ for {\mfs} and $x_{\mathrm{Co}} > 0.05$ for {\mcs}) accompanied by a suppression of $T_{2}$, the cross-over temperature to the FD regime, above $x_{\mathrm{Fe}} = 0.22$~\cite{2010:Pfleiderer:JPhysCondensMatter, 2010:Bauer:PhysRevB}. This conjecture was corroborated by neutron spin echo measurements of the line-width of quasi-elastic scattering for $x_{\mathrm{Fe}} = 0.08$ when approaching $T_{c}$, and susceptibility data for $x_{\mathrm{Fe}} = 0.12$~\cite{2011:Grigoriev:PhysRevB}. Moreover, careful comparison of the effects of cobalt doping in {\mcs} with {\mfs} revealed remarkably similar behavior. This provided striking evidence that the number of valence electrons controls the properties~\cite{2010:Bauer:PhysRevB, 2010:Teyssier:PhysRevB, 2017:Dhital:PRB}. In turn, the electronic structure and concomitant magnetic properties of {\mfs} and {\mcs} are surprisingly similar.

In recent years several studies have addressed the nature of the fluctuations in {\mfs} at large doping levels~\cite{2014:DemJETP, 2015:Glushkov:PRL, 2016:Demishev:aaa}. Unfortunately, these studies infer their conclusions from data recorded at large magnetic fields, where the effects of MCAs no longer dominate. Likewise, the emergence of a Griffiths' regime with short-range order driven by the inherent disorder in {\mfs} at large doping levels was inferred from large magnetic fields, where the role of disorder at zero and low magnetic fields cannot be assessed. 

In a different set of studies, magnetization, susceptibility and SANS were reported for {\mfs} up to $x=0.32$ \cite{2018:Bannenberg:Mag,2018:Bannenberg:SANS}. The magnetization and susceptibility confirmed earlier work, lacking information on anisotropies. Further, the SANS measurements claim to observe a change of the direction of zero-field-cooled helimagnetic order from $\langle 111\rangle$ to $\langle 110 \rangle$ without offering any theoretical explanation whatsoever \cite{2018:Bannenberg:SANS}. Moreover, data reported in this paper was recorded for one crystallographic orientation only without further information on the full three-dimensional intensity distribution. In addition, the zero-field cooled data exhibit single domain populations suggesting either strained samples or the presence of a parasitic field, and data were not properly centered shedding doubts as to the correct determination of the crystallographic direction of the intensities.

Important insights are here provided by the electrical transport properties, where a large increase of the residual resistivity $\rho_{0}$ raises concerns on the detailed analysis of the temperature dependent part of the resistivity, $\Delta\rho$, and a possible NFL dependence~\cite{2009:Meingast:Book, 2014:Franz:PhysRevLett}. Moreover, the magnetic properties observed in the magnetization and ac susceptibility appeared to become more isotropic with increasing $x$, where the effects of disorder as opposed to MCAs could not be assessed for lack of microscopic information. In turn, an important unresolved question concerns to what extent the disorder causes a smearing of thermodynamic transitions in the magnetic phase diagram, thus possibly dominating the properties at scales much higher than the MCAs.


\subsection{Microscopic origin of interactions}
\label{microsc}

A long history of research on the nature of the magnetic properties of MnSi together with recent work in {\mfs} and {\mcs} connect the hierarchical energy scales in these materials with the framework of itinerant-electron magnetism and the Fermi surface.
In the history of itinerant-electron magnetism, the properties of MnSi have been of central importance in their own right. In the field-polarized state of MnSi the ordered magnetic moment is fully accounted for by a weakly exchange-split Fermi surface~\cite{1986:Taillefer:JMMM, 1987:Lonzarich:JMMM, 1988:Lonzarich:JMMM}. The underlying exchange-enhanced Fermi liquid ground state forms the starting point for a Ginzburg--Landau model of the thermodynamic properties. Taking into account the measured spectrum of spin fluctuations self-consistently, this model has provided a remarkably successful quantitative account of the magnetic equation of state and associated properties such as the transition temperature~\cite{1985:Lonzarich:JPCSP, 1985:Moriya:book}. 

Starting from these well-understood electronic properties of MnSi, recent measurements of the ordinary, anomalous and topological Hall effect in {\mfs} and {\mcs} could be explained by ab initio calculations of a rigidly exchange-split Fermi surface~\cite{2012:Schulz:NaturePhys, 2014:Franz:PhysRevLett}. This provided very sensitive indirect evidence that the details of the Fermi surface and Berry curvature under doping are correctly predicted by density functional theory. It established, in particular, that the effects of spin--orbit coupling in the Fermi surface are captured satisfactorily. As part of this study, various numerical tests established that defect related scattering on different Fermi surface sheets is not relevant for the overall understanding of the Hall effect. Moreover, compositional studies of {\mcs} and {\mis} exploring the role of the valence electron count for the electronic structure confirm the suitability of rigid exchange splitting in the description of the electronic structure~\cite{2017:Dhital:PRB}. 

A seminal ab initio study attributed the strength of the DMI in MnSi quantitatively to the Berry curvature of the electronic structure~\cite{2013:Freimuth:PRB,2014:Freimuth:JPCM} consistent with the Hall effect \cite{2014:Franz:PhysRevLett}. An important test is here the consistency of changes of $\vert Q\vert$ under Fe and Co doping, not attempted before. Last but not least, numerous studies of weak itinerant electron magnets attribute MCAs to a small number of electronic states in the neighborhood of degeneracies in the band structure close to Fermi energy \cite{sigfusson:82,kondorsky:74,2008:Buenmann:PhysRevLett}.

Taken together, the connection between the hierarchy of energy scales and their evolution under doping may be summarized as follows. First, the ferromagnetic exchange coupling, representing the strongest scale, is associated with itinerant-electron ferromagnetism and thus the exchange splitting. Viewed in terms of the Stoner criterium of itinerant magnetism this translates into the density of states at the Fermi level. Second, the DMI as the intermediate scale which controls the magnitude of twisting, $\vert Q\vert$, reflects the Berry curvature of the electronic structure. For rigid band splitting under doping only moderate changes may be expected. Third, band degeneracies close to the Fermi level are likely to control the MCAs.

This provides the starting point, when discussing the effect of doping in {\mfs} and {\mcs} and the plausibility of a shift of the relative importance of the character of the MCAs from terms that are fourth to terms the sixth order in spin-orbit coupling as our main conclusion, while the strength of the combined effect of defects and disorder with the MCAs varies only weakly.


\section{Experimental Methods}
\label{ExpMeth}

\subsection{Sample preparation}
\label{samples}

In our study we investigated large single crystals of MnSi, Mn$_{1-x}$Fe$_{x}$Si with $x_{\mathrm{Fe}} = 0.02$, 0.04, 0.06, 0.08, 0.10, 0.12, 0.16, 0.19, 0.22, and Mn$_{1-x}$Co$_{x}$Si with $x_{\mathrm{Co}} = 0.02$, 0.04 grown by means of the optical floating-zone technique under ultra-high vacuum compatible conditions~\cite{2011:Neubauer:RevSciInstrum, 2016:Bauer:RevSciInstrum, 2016:Bauer:RevSciInstrum2}. The growth process and the metallurgical characterization used for all samples followed the procedure described in Ref.~\onlinecite{2010:Bauer:PhysRevB}. Data recorded in some of the samples had been reported in Ref.~\onlinecite{2010:Bauer:PhysRevB}. We wish to emphasize, that all data reported in this paper were measured specifically for the present study and were not reported elsewhere before. The only exception are the data shown in Fig.~\ref{figure01}(a2) for pure MnSi, which had been reported in Ref.~\onlinecite{2017:Bauer:PRB}, and selected data points in the magnetic phase diagrams in Figs.~\ref{figure01} and \ref{figure03}, which had been reported in Ref.~\onlinecite{2010:Bauer:PhysRevB}. These data are included in the presentation to permit convenient comparison with previous work.

The following empirical observations establish excellent compositional homogeneity and the absence of grown-in lattice strain for the entire series of compositions. Namely, measurements of the physical properties of samples cut from the start and the end of the float-zoned ingots displayed identical magnetic properties within experimental accuracy, establishing excellent homogeneity. Moreover, optical float-zoning of seed and feed rods with differing compositions allowed to prepare single crystals with compositional gradients, where the distance of the compositional change along the growth direction corresponded to the length of the molten zone (for details we refer to Ref.~\onlinecite{Neubauer:PhD-thesis}). This implies fast diffusion rates of the dopants, underscoring the high homogeneity inferred from the physical properties. 

Further, using oriented seed crystals, large single crystals of pure MnSi with diameters up to 10~mm were prepared for different studies, where the growth directions were chosen to be along $\langle100\rangle$, $\langle110\rangle$, $\langle111\rangle$, and $\langle211\rangle$. On a similar note, carefully inspecting the orientation of the Mn$_{1-x}$Fe$_{x}$Si and Mn$_{1-x}$Co$_{x}$Si single crystals, all of which were prepared using polycrystalline seeds, there was no evidence suggesting a trend for preferred growth directions either. Finally, the distribution of the SANS intensity patterns recorded under zero-field cooling were the same for equivalent crystallographic directions as reported below. This contrasts the effects of uniaxial stress, which have been found to impose changes of domain populations and preferred propagation directions of the magnetic modulation~\cite{2015:Chacon:PhysRevLett, 2015:Nii:NatCommun, 2017:Seki:PRB}. Hence, there was no evidence suggesting grown-in strain, as may be expected from temperature or compositional gradients during growth. 

For the small-angle neutron scattering studies cylinder-shaped single-crystal samples with a diameter of 6~mm and a height of typically 10~mm--20~mm were prepared using a wire saw. The orientation of the samples was determined by means of x-ray Laue diffraction. For the measurements of the bulk properties discs of 1~mm or 2~mm thickness were cut from the same ingots, where the direction normal to the disc was oriented along a $\langle110\rangle$ axis. 

The magnetic susceptibility and specific heat of MnSi were measured on two cubic samples prepared from the same disc, where the edges of the cubes had a length of 2~mm. One cube was oriented such that two surfaces were perpendicular to $\langle100\rangle$ and four surfaces were perpendicular to $\langle110\rangle$. The second cube was oriented such that two surfaces each were oriented perpendicular to $\langle110\rangle$, $\langle111\rangle$, and $\langle211\rangle$. 

For measurements of the magnetic susceptibility of Mn$_{1-x}$Fe$_{x}$Si and Mn$_{1-x}$Co$_{x}$Si cubic samples were prepared, where the edges had a length of 1~mm. These samples were oriented such that two surfaces were perpendicular to $\langle100\rangle$ and four surfaces were perpendicular to $\langle110\rangle$. The specific heat of Mn$_{1-x}$Fe$_{x}$Si and Mn$_{1-x}$Co$_{x}$Si was measured in samples representing a quarter of a disc with a radius of 3~mm and a height of 1~mm cut from the same discs used for the preparation of the cubic samples. For the specific heat measurements the magnetic field was applied along the axes normal to the discs corresponding to $\langle110\rangle$. 

\subsection{Bulk properties}

The magnetization, ac susceptibility, and specific heat were measured in a Quantum Design physical property measurement system~(PPMS). The magnetization was determined with an extraction technique. The ac susceptibility was measured at an excitation amplitude of 1~mT and an excitation frequency of 911~Hz~\cite{2012:Bauer:PhysRevB}. In comparison to magnetization and ac susceptibility data reported in Ref.~\onlinecite{2010:Bauer:PhysRevB}, data for the study reported here were recorded at a much higher density with improved parameter settings to permit detailed comparison of $\mathrm{Re}\,\chi_{\mathrm{ac}}$ and $\mathrm{d}M/\mathrm{d}H$ as well as the properties for different crystallographic orientations. Such a comparison was not possible with data reported in Ref.~\onlinecite{2010:Bauer:PhysRevB}. As mentioned above the susceptibility data as a function of field for pure MnSi (cf.\ Fig.~\ref{figure01}(a2)) have been shown in Ref.~\onlinecite{2017:Bauer:PRB} before. None of the other susceptibility data shown in this paper were reported before. 

The specific heat of MnSi and Mn$_{1-x}$Fe$_{x}$Si for compositions up to $x_{\mathrm{Fe}} = 0.06$ was determined by means of a quasi-adiabatic large heat pulse method where heat pulses had a typical size of 30\% of the temperature of the bath~\cite{2013:Bauer:PhysRevLett}. For Mn$_{1-x}$Fe$_{x}$Si with $x_{\mathrm{Fe}} = 0.08$ and $x_{\mathrm{Fe}} = 0.10$ in addition a conventional small heat pulse method was used, where typical heat pulses were around 1\% of the temperature at the start of the pulse. The specific heat data shown below, which focus on the thermodynamic anomalies associated with the skyrmion lattice phase in Mn$_{1-x}$Fe$_{x}$Si, have not been reported before. In particular, the data reported in Ref.~\onlinecite{2010:Bauer:PhysRevB} did not allow to discern the anomalies of the skyrmion lattice phase. Likewise, in order to compare data for different field direction under unchanged demagnetization conditions (not shown, cf.\ Ref.~\onlinecite{2016:Bauer:Book} for data with field parallel to $\langle100\rangle$), the specific heat of pure MnSi was measured on cube-shaped samples, in contrast to Ref.~\onlinecite{2013:Bauer:PhysRevLett}.

All experimental data are shown as a function of applied magnetic field without correction for demagnetizing fields, whereas the magnetic phase diagrams inferred from these data are shown as a function of internal field. The demagnetizing factors of the samples were determined by approximating the sample shapes as rectangular prisms~\cite{1998:Aharoni:JApplPhys}.

For what follows, it is helpful to introduce three temperature versus field histories. Zero-field cooling~(ZFC) refers to a protocol, where the sample is initially cooled down from a high temperature in zero magnetic field and data recorded either as a function of magnetic field, or as a function of temperature once the desired field value had been set. In contrast, field cooling down~(FCD) refers to a protocol, where data are recorded while cooling the sample from a high temperature in a fixed magnetic field. High-field cooling~(HFC) refers, finally, to a protocol, where the sample is initially cooled down from a high temperature in a magnetic field larger than $H_{c2}$ (the value of the conical-to-field-polarized transition). Data are then recorded either as a function of magnetic field or as a function of temperature once the desired field value had been set. 

Keeping in mind the different temperature versus field protocols, data recorded as a function of magnetic field presented in this paper are denoted as follows: (1)~increasing field after zero-field cooling to the desired temperature, (2)~decreasing field starting in the field-polarized state at $+1~\mathrm{T}$, and (3)~increasing field starting in the field-polarized state at $-1~\mathrm{T}$. Sweeps (2) and (3) correspond to HFC conditions but differ in terms of the sweep direction.


\subsection{Small-angle neutron scattering}

The small-angle neutron scattering measurements were carried out on the beamlines RESEDA~\cite{2007:Haussler:PhysicaB, 2015:Franz:JLSFR} and MIRA~\cite{2015:Georgii:JLSFR} at the Heinz Maier-Leibnitz Zentrum~(MLZ) and the beamline SANS-II at the Paul Scherrer Institut~(PSI). 

At MIRA and RESEDA neutrons were used with a wavelength $\lambda = 11.5~\textrm{\AA}\pm5\%$ and $\lambda = 5.42~\textrm{\AA}\pm10\%$, respectively. On both beamlines neutrons were detected by means of a CASCADE detector~\cite{2011:Haussler:RevSciInstrum} with an active area of $200\times200~\mathrm{mm}^{2}$ ($128\times128$ pixels). For the experiments at MLZ samples were cooled down to 3.5~K with a standard pulse tube cooler, whereas a $^{3}$He insert and a dilution insert were used for studies at temperatures down to 0.5~K and 50~mK, respectively. Moreover, magnetic fields up to 300~mT were generated with a bespoke set of water-cooled Helmholtz coils. At SANS-II neutrons with a wavelength $\lambda = 4.9~\textrm{\AA}\pm10\%$ were used, where a $^{4}$He cryostat with a superconducting magnet provided temperatures down to 3~K under horizontal fields up to 11~T.

In order to track the precise details of the SANS intensity distribution comprising sharp spots and smeared out features, data were recorded for 40 different orientations of the incoming neutron beam covering a sample rotation of $180^{\circ}$. For the experiments the set-up of sample, cryostat, and magnet were carefully aligned, where the samples was oriented such that a $\langle110\rangle$ axis corresponded to the vertical axis. During rocking scans the sample and magnet were rotated together, where $\omega$ denotes the rocking angle. Scattering patterns shown in the following represent sums over rocking scans. These patterns are dominated by intensity on a ring in reciprocal space. In turn, the intensity distribution was evaluated as a function of the azimuthal angle $\alpha$, where $\alpha = 0$ corresponds to the vertical direction. The radius of the ring, denoted $Q_h$, corresponds to the size of the helical modulation vector, which varies as a function of composition as described below. 

Unfortunately, quantitative comparison of the neutron intensities between different samples proves to be rather difficult, since the measurements were carried out at three different beamlines and for samples of different shape and volume. However, as most samples were studied at RESEDA under comparable conditions, it is at least possible to note that the integrated intensities are consistent with the decrease of the square of the ordered moment as a function of increasing $x$~\cite{2010:Bauer:PhysRevB}.


\section{Bulk properties}
\label{ResultsBulk}

The presentation of the bulk properties focuses on those aspects relevant for the microscopic evolution of the spontaneous magnetic order in {\mfs} and {\mcs} pursued in the present paper. This concerns the magnetic phase diagrams, putative evidence for the relevance of domain populations and disorder-induced pinning, thermodynamic signatures of phase boundaries, and evidence of magnetic anisotropies. The account begins in Sec.~\ref{mfs-susc} with an overview of the magnetic phase diagrams of {\mfs}, the underlying ac susceptibility, and the susceptibility calculated from the magnetization, continues in Sec.~\ref{mfs-specheat} with the specific heat of {\mfs} in the regime of the skyrmion lattice phase, and concludes in Sec.~\ref{mcs-susc} with the phase diagrams and susceptibilities of {\mcs}. 


\subsection{Susceptibility and magnetic phase diagrams of Mn$_{1-x}$Fe$_{x}$Si}
\label{mfs-susc}

\begin{figure*}
\includegraphics[width=1.0\linewidth]{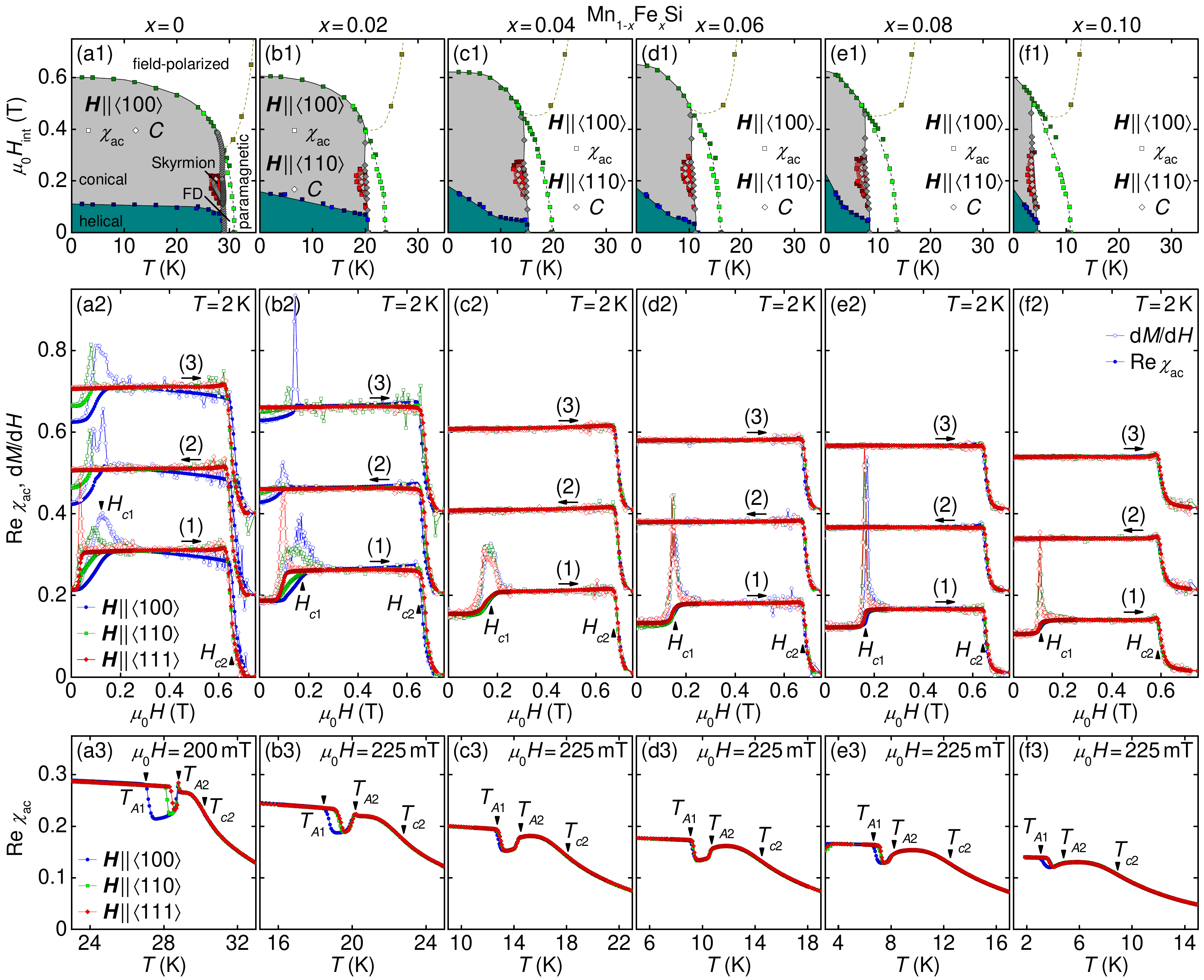}
\caption{Magnetic phase diagrams, ac susceptibility, $\mathrm{Re}\,\chi_{\mathrm{ac}}$, and susceptibility calculated from the magnetization, $\mathrm{d}M/\mathrm{d}H$, of Mn$_{1-x}$Fe$_{x}$Si. \mbox{(a1)--(f1)}~Magnetic phase diagrams as inferred from the susceptibilities and specific heat data for MnSi and Mn$_{1-x}$Fe$_{x}$Si up to $x_{\mathrm{Fe}} = 0.10$. Field values have been corrected for demagnetization effects. \mbox{(a2)--(f2)}~Comparison of $\mathrm{Re}\,\chi_{\mathrm{ac}}$ and $\mathrm{d}M/\mathrm{d}H$ for magnetic field along the major crystallographic axes at low temperatures. Notation refers to the following types of field sweeps: (1)~increasing field following initial zero-field cooling, (2)~decreasing field starting at $+1~\mathrm{T}$, and (3)~increasing field starting at $-1~\mathrm{T}$. Data are shifted vertically by 0.2 between (1), (2), and (3) for better visibility. \mbox{(a3)--(f3)}~Susceptibility as a function of temperature at an applied magnetic field selected such that the skyrmion lattice phase displays the largest temperature range. Data are shown for field along the three major crystallographic axes. Transition fields and temperatures values are marked for $\bm{H}\parallel\langle100\rangle$ after zero-field cooling only, for better visibility.}
\label{figure01}
\end{figure*}

Shown in Fig.~\ref{figure01}(a1) is the magnetic phase diagram of MnSi in excellent agreement with Refs.~\onlinecite{2012:Bauer:PhysRevB, 2013:Bauer:PhysRevLett}. The transition temperatures and transition fields shown here were inferred from the susceptibility (squares) and the specific heat (diamonds) for field applied along a crystallographic $\langle100\rangle$ direction after zero-field cooling. Six regimes may be distinguished: (i)~ multi-domain helical order at low magnetic fields, (ii)~conical order at low temperatures under moderate applied fields, (iii)~the field-polarized state for low temperatures and high fields, (iv)~skyrmion lattice order under small applied fields within the conical state just below $T_{c}$, (v)~the fluctuation-disordered~(FD) regime of the paramagnetic state just above $T_{c}$, and (vi)~the paramagnetic state at high temperatures and low fields. 

As a function of magnetic field the susceptibility of MnSi, shown in Fig.~\ref{figure01}(a2), increases across the helix reorientation at $H_{c1}$, followed by a strong decrease at the transition from the conical to the field-polarized state. Whereas the susceptibility calculated from the measured magnetization, $\mathrm{d}M/\mathrm{d}H$ (open symbols), displays a pronounced peak at $H_{c1}$, there is no such peak in the real part of the ac susceptibility, $\mathrm{Re}\chi_{\mathrm{ac}}$ (solid symbols). This discrepancy reflects the difference of the fast response of single spins probed by the ac susceptibility as opposed to the full response of the system tracked in the magnetization, which includes also slow changes of the propagation vector of entire helical domains~\cite{2012:Bauer:PhysRevB, 2017:Bauer:PRB}. In contrast, $\mathrm{d}M/\mathrm{d}H$ tracks $\mathrm{Re}\chi_{\mathrm{ac}}$ accurately at $H_{c2}$, the transition between the conical and the field-polarized state. 

Further, in the temperature dependence of the susceptibility of MnSi, shown in Fig.~\ref{figure01}(a3), the regime of the skyrmion lattice phase corresponds to a reduced value between $T_{A1}$ and $T_{A2}$. Above $T_{A2}$ the fluctuation-disordered regime is entered where a point of inflection at $T_{c2} > T_{A2}$ marks the crossover to the paramagnetic properties at high temperatures. We follow the convention introduced in Refs.~\onlinecite{2010:Bauer:PhysRevB, 2012:Bauer:PhysRevB} and denote the point of inflection at finite field by $T_{c2}$ and at zero field by $T_{2}$, i.e., $T_{c2}(H\to0) = T_{2}$. Note that, as for the helix reorientation, magnetic field sweeps in the regime of the skyrmion lattice phase display a pronounced peak in $\mathrm{d}M/\mathrm{d}H$ that is not tracked in $\mathrm{Re}\,\chi_{\mathrm{ac}}$ (not shown)~\cite{2012:Bauer:PhysRevB}.

The magnetic phase diagram of MnSi displays several distinct changes as a function of crystallographic orientation, all of which may be fully accounted for in terms of the cubic MCAs. Namely, the field value of the reorientation between the helical and the conical state, denoted $H_{c1}$, changes strongly with crystallographic orientation but is essentially constant as a function of temperature for each given orientation. Further, subject to field direction the character of the phase transformation at $H_{c1}$ may be either that of a first-order transition, a second-order transition, or a crossover~\cite{2017:Bauer:PRB}. In turn, depending on the temperature versus field protocol, signatures of the susceptibility at $H_{c1}$ differ due to differences of the occupation of the helical domains. In comparison, the transition field between the conical and the field-polarized state at $H_{c2}$ as a function orientation varies by a few percent only. Finally, the field range of the skyrmion lattice phase is essentially insensitive to crystallographic orientation whereas the temperature range varies by a factor of two. 

As a function of increasing iron content the qualitative appearance of the magnetic phase diagram of {\mfs} for $x_{\mathrm{Fe}} \leq 0.1$ remains essentially unchanged as shown in Figs.~\ref{figure01}(b1) to \ref{figure01}(f1). The diagrams are in excellent agreement with the literature~\cite{2009:Grigoriev:PhysRevB, 2010:Bauer:PhysRevB, 2010:Pfleiderer:JPhysCondensMatter,2018:Bannenberg:Mag}, representing updated versions where we have taken into account small changes as explained below based on the high resolution of our specific heat data. In view of the isotropic behavior observed in the susceptibility presented below, the phase diagrams shown here for {\mfs} combine data points inferred from the susceptibility for field parallel to $\langle100\rangle$ and the specific heat for field parallel to $\langle110\rangle$ in the same diagrams. Typical susceptibility data as a function of magnetic field recorded at 2~K are shown in Figs.~\ref{figure01}(b2)--\ref{figure01}(f2), where magnetic field was applied along $\langle100\rangle$ (blue circles), $\langle110\rangle$ (green squares), and $\langle111\rangle$ (red diamonds). The analogous temperature dependence at a fixed applied magnetic field recorded after zero-field cooling are shown in Figs.~\ref{figure01}(b3) to \ref{figure01}(f3).

With increasing iron content the upper critical field $H_{c2}$ is essentially unchanged, while the transition temperatures decrease consistent with previous reports. The field value of the helix reorientation at $H_{c1}$, as determined after zero-field cooling, features a small increase of its extrapolated zero-temperature value by a factor of two, and a change of its temperature dependence. 

As a function of the magnetic field, the susceptibilities for all compositions $x_{\mathrm{Fe}} \geq 0.04$ display the same characteristics that are reminiscent of MnSi but differ in details as elaborated in the following. The properties for $x_{\mathrm{Fe}} = 0.02$ are intermediate between pure MnSi and the samples with $x_{\mathrm{Fe}} \geq 0.04$. Namely, following zero-field cooling, denoted (1), the ac susceptibility exhibits an increase at the helix orientation at $H_{c1}$, followed by a pronounced drop at the conical-to-field-polarized transition $H_{c2}$. The transition at $H_{c1}$ is accompanied by a pronounced maximum in $\mathrm{d}M/\mathrm{d}H$, whereas no peak is seen in $\mathrm{Re}\,\chi_{\mathrm{ac}}$ for the excitation frequency of 911~Hz. Moreover, $\mathrm{d}M/\mathrm{d}H$ and $\mathrm{Re}\,\chi_{\mathrm{ac}}$ track each other accurately across $H_{c2}$.

However, in comparison with MnSi careful inspection of $\mathrm{d}M/\mathrm{d}H$ and $\mathrm{Re}\chi_{\mathrm{ac}}$ reveals several important differences as a function of crystallographic orientation and field versus temperature protocols. Namely, in stark contrast to MnSi, the anomalies at $H_{c1}$ do not show any dependence on crystallographic orientation and the value of $H_{c1}$ as extrapolated to zero temperature increases weakly by a factor of two when going above $x_{\rm Fe}\approx 0.04$. Moreover, with increasing temperature $H_{c1}$ decreases strongly.
All of these characteristics may be explained in terms of defect-induced pinning and thermally driven unpinning, where the effects of defects and disorder are enhanced in comparison to MCAs. 

This observation is corroborated by the behavior in field sweeps starting in the field-polarized state, denoted (2) and (3). While the field-polarized-to-conical transition at $H_{c2}$ is unchanged, all signatures of the helix reorientation at $H_{c1}$ are absent. Indeed, as presented in Sec.~\ref{ResultsSANS}, once the conical state has been formed the pristine, multi-domain helical state is not recovered under decreasing field strength, even when sweeping through zero magnetic field [curve (3)]. Instead, the pinning appears to favor a single-domain state parallel to the field, without signature of the transition at $H_{c1}$ in the magnetization and ac susceptibility as reported for pure MnSi under field along $\langle 111\rangle$ \cite{2013:Narozhnyi:JETP,2015:Narozhnyi:PRB,2017:Bauer:PRB}.

It is interesting to note, that identical differences as compared with MnSi have also been reported for the helix reorientation in Fe$_{1-x}$Co$_{x}$Si~\cite{2010:Munzer:PhysRevB, 2016:Bauer:PhysRevB}, namely isotropic values of $H_{c1}$, a strong decrease with increasing temperature, and lack of recovery of the helical state once the conical state has been induced. This suggests that structural disorder and defects as compared with the strength of MCAs dominates the properties of the doped cubic chiral magnets. We return to this question when presenting our neutron scattering results in Sec.~\ref{ResultsSANS}, which provide clear evidence of defect-induced domain pinning and that leading-order contributions to the cubic MCAs decrease with increasing iron or cobalt concentration.

Similar changes of characteristics as compared to those at $H_{c1}$ are also observed in the temperature dependence of the ac susceptibility under an applied field after zero-field cooling shown in Figs.~\ref{figure01}(a3) to \ref{figure01}(f3). For the data presented here, the field value was chosen such that the skyrmion lattice phase covers the largest possible temperature range. In MnSi, the skyrmion lattice displays the largest extent as a function of temperature for field along $\langle100\rangle$, while the phase pocket is smallest for field along $\langle111\rangle$, cf.\ Fig.~\ref{figure01}(a3). This behavior is consistent with easy $\langle111\rangle$ axes for the helical modulation. With increasing $x_{\mathrm{Fe}}$, the anisotropy vanishes for $x_{\mathrm{Fe}} \geq 0.04$ as shown in Figs.~\ref{figure01}(b3) to \ref{figure01}(f3). For $x_{\mathrm{Fe}} \geq 0.08$, the temperature range for field along $\langle100\rangle$ again becomes slightly larger compared to the other field directions, characteristic of slightly stronger anisotropies. In addition, with increasing $x_{\mathrm{Fe}}$ the signature at $T_{A2}$ becomes increasingly smeared out, while the temperature width of the fluctuation-disordered regime increases both in absolute terms and in units of $T_{c}$~\cite{2009:Grigoriev:PhysRevB}. Under field cooling the extent of the skyrmion lattice phase is unchanged (not shown).

Taken together, our susceptibility data establish that the helix reorientation and the temperature range of the skyrmion lattice state in {\mfs} become essentially isotropic for $x_{\mathrm{Fe}} \geq 0.04$. At the same time the signatures of the skyrmion lattice phase are still clearly visible in the susceptibility. In contrast, once the conical state has been stabilized under applied magnetic field, the anisotropies are too weak to overcome the local pinning of the helices due to defects and disorder. As a result, the helices remain aligned along the field direction characteristic of a single-domain state without signatures of a helix reorientation transition in the magnetization and susceptibility. This is consistent with the SANS data presented in Sec.~\ref{ResultsSANS} below.


\subsection{Specific heat of Mn$_{1-x}$Fe$_{x}$Si}
\label{mfs-specheat}

\begin{figure}
\includegraphics[width=1.0\linewidth]{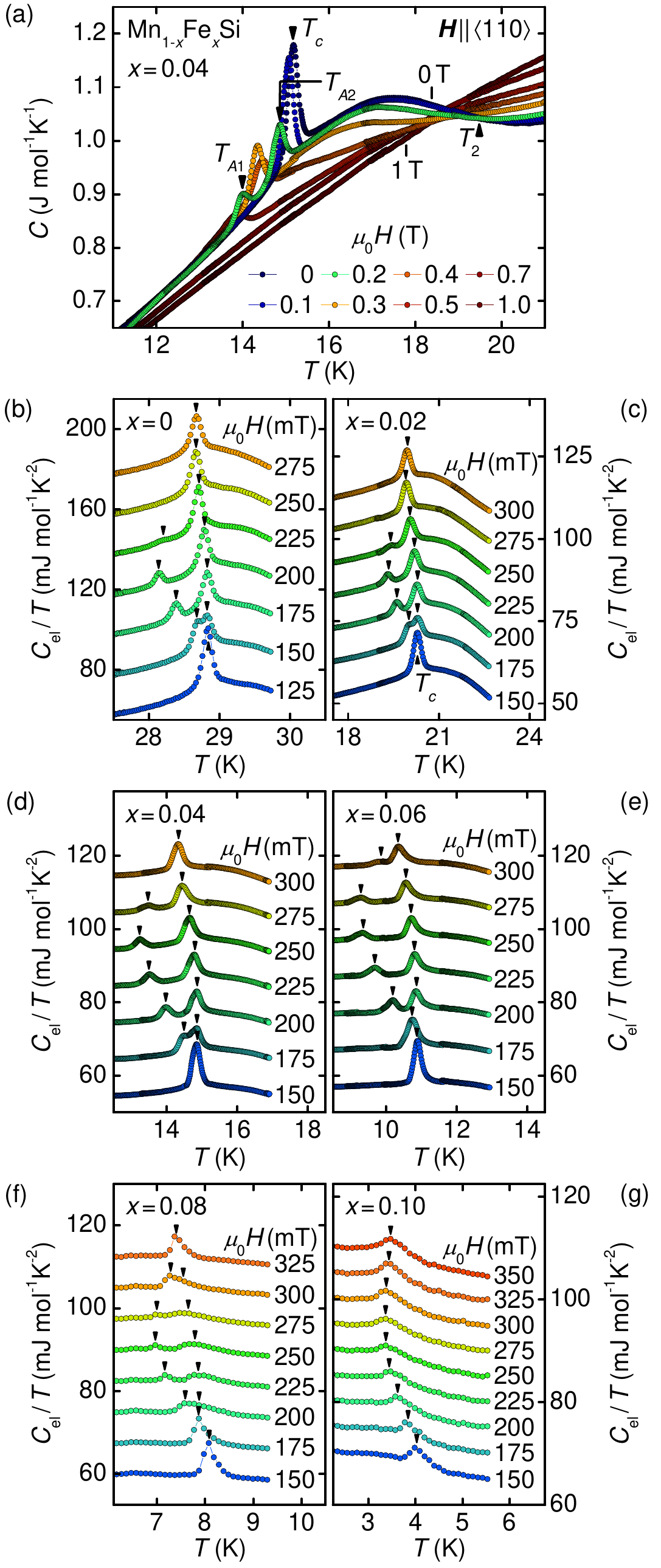}
\caption{Specific heat of Mn$_{1-x}$Fe$_{x}$Si for magnetic field parallel to $\langle110\rangle$. (a)~Specific heat as a function of temperature for $x_{\mathrm{Fe}} = 0.04$ and fields up to 1~T. \mbox{(b)--(g)}~Electronic contribution to the specific heat divided by temperature, $C_{\mathrm{el}}/T$, for MnSi and Mn$_{1-x}$Fe$_{x}$Si up to $x_{\mathrm{Fe}} = 0.10$ and fields values around the skyrmion lattice pocket. Arrows mark the transition temperatures. Data have been shifted vertically for clarity.}
\label{figure02}
\end{figure}

The evidence for the effects of disorder and defects under iron doping seen in the susceptibility at $H_{c1}$ and the temperature range of the skyrmion lattice raise the question, to what extent the different ordered states are still separated by well-defined thermodynamic phase transitions. Namely, are the effects of defects and disorder weaker then the ferromagnetic exchange and DMI, which drive the formation of long-range order. Consistent with the literature, it was not possible to resolve anomalies in the specific heat at $H_{c1}$.  We focus therefore on the skyrmion lattice phase.

Shown in Fig.~\ref{figure02}(a) is the specific heat for Mn$_{1-x}$Fe$_{x}$Si with $x = 0.04$ and fields up to 1~T applied along $\langle110\rangle$. For the large pulse method used here the specific heat displays much higher resolution as compared with previous studies~\cite{2010:Bauer:PhysRevB}. Data for {\mfs} up to high values of $x_{\mathrm{Fe}}$ are highly reminiscent of MnSi~\cite{2013:Bauer:PhysRevLett}. As reported before, for zero magnetic field the peak at the onset of helical order at $T_{c}$ is consistent with the fluctuation-induced first-order transition observed in MnSi~\cite{2010:Bauer:PhysRevB, 2013:Janoschek:PhysRevB}. The transition temperature is in excellent agreement with the susceptibility and small-angle neutron scattering. The peak resides on top of a broad shoulder characterized by an invariant crossing point for small fields at $T_{2}$ that coincides with a point of inflection in the susceptibility, i.e., the behavior is consistent with a Vollhardt invariance~\cite{2010:Bauer:PhysRevB, 1997:Vollhardt:PhysRevLett}. 

For larger magnetic fields the peak in the specific heat assumes the temperature dependence of a lambda anomaly characteristic of a second-order mean-field transition. The evidence for such a field-induced tricritical point in Mn$_{1-x}$Fe$_{x}$Si, akin to pure MnSi, implies that the crossover line between the paramagnetic and the field-polarized regime emanates from this point. Based on this conjecture, the position of the lines shown in the magnetic phase diagrams in Fig.~\ref{figure01} as a guide to the eye were adjusted as compared with those presented in Ref.~\onlinecite{2010:Bauer:PhysRevB}.

The properties of a putative Brazovskii transition under compositional doping are corroborated by the temperature dependence of the susceptibility, which displays a point of inflection at $T_{2}$, consistent with the isotropic SANS intensity distributions observed just above $T_{c}$~\cite{2009:Grigoriev:PhysRevB}. Here the combination of qualitatively unchanged magnetic phase diagrams with the clear thermodynamic signatures of the skyrmion lattice phase suggests strongly that {\mfs} remains metallurgically homogeneous under iron doping and may be described by a combination of a hierarchy of energy scales, where defect- and disorder-induced pinning exceeds the MCAs.

The much higher resolution of our specific heat data allows to resolve, to the best of our knowledge for the first time in {\mfs}, the presence of two clearly discernible peaks in $C(T)$ for magnetic fields around 0.2~T, cf.\ Fig.~\ref{figure02}(a). Both peaks are unambiguously connected with the phase transitions of the skyrmion lattice state at $T_{A1}$ and $T_{A2}$. Providing unambiguous evidence of well-defined thermodynamic phase transitions of the skyrmion lattice phase under iron doping all the way up to $x_{\mathrm{Fe}} = 0.08$, the corresponding data are shown in Figs.~\ref{figure02}(b) to \ref{figure02}(g) in terms of the electronic contribution to the specific heat divided by temperature, $C_{\mathrm{el}}/T$. Here lattice contributions were subtracted corresponding to a Debye temperature $\mathit{\Theta}_{\mathrm{D}} = 513$~K that is essentially unchanged as a function of iron doping~\cite{2010:Bauer:PhysRevB}. For $x_{\mathrm{Fe}} = 0.10$ only the peak at $T_{A1}$ is resolved, consistent with the rounding of the corresponding signatures in the susceptibility. 

Thus, the following observations reflect the role of defects and pinning under doping: First, as presented in Sec.\,\ref{mfs-susc}, variations of the temperature range of the skyrmion lattice phase as a function of crystallographic orientation, which reflect the effects of MCAs, vanish under doping.  Second, the absolute field range of the skyrmion lattice phase remains essentially unchanged. Third, the phase transition between the conical and the skyrmion lattice phase is accompanied by anomalies in the specific heat. Taken together this underscores, that the effects of pinning are similar in strength to the MCAs. At the same time the strength of the MCAs may be changing only weakly. Yet, the skyrmion lattice phase is thermodynamically well-defined despite of the effects of pinning and disorder which must still be weak as compared with the exchange interactions and DMI.


\subsection{Susceptibility and magnetic phase diagrams of Mn$_{1-x}$Co$_{x}$Si}
\label{mcs-susc}

\begin{figure}
\includegraphics[width=1.0\linewidth]{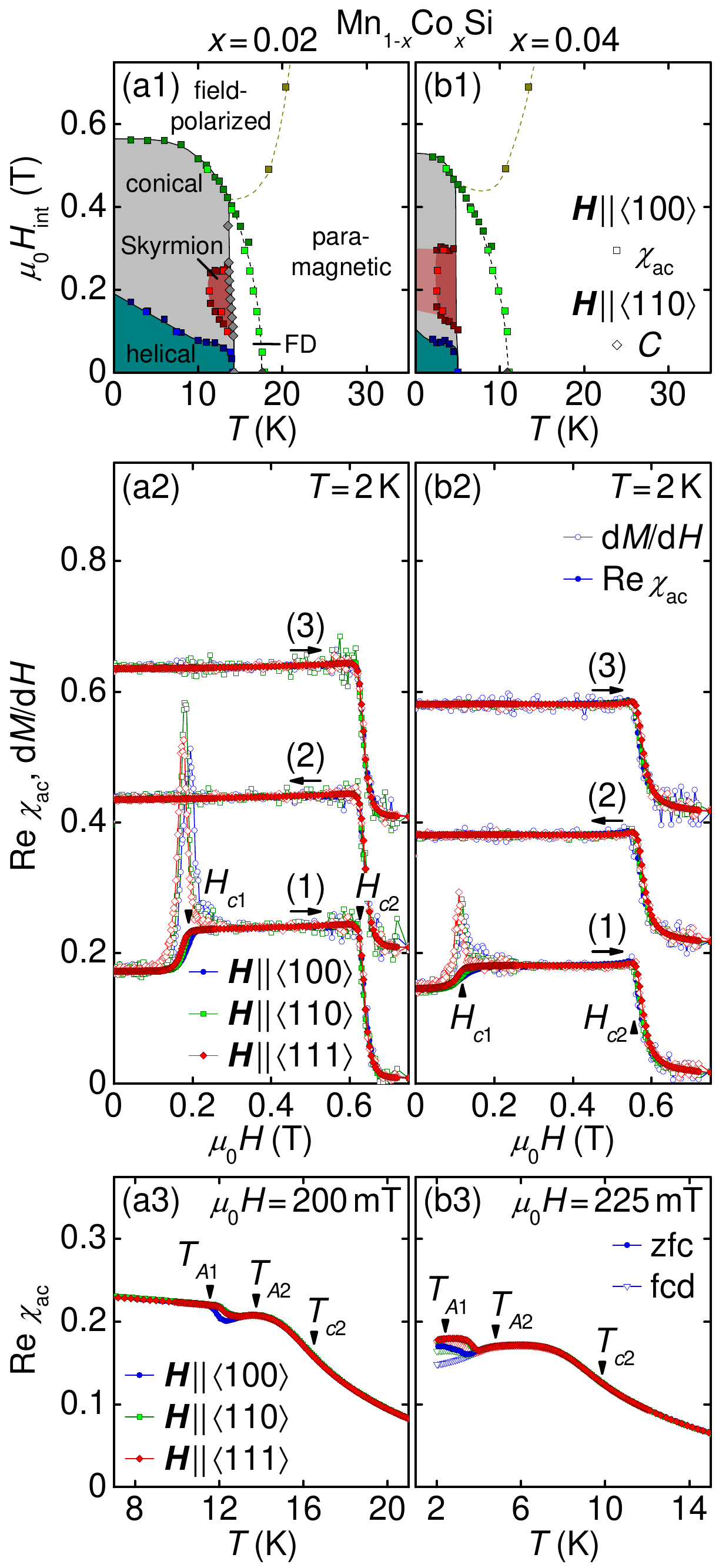}
\caption{Magnetic phase diagrams and susceptibility data for Mn$_{1-x}$Co$_{x}$Si. \mbox{(a1)--(b1)}~Magnetic phase diagrams as inferred from susceptibility and specific heat data. \mbox{(a2)--(b2)}~Susceptibility as a function of field along the major crystallographic axes at low temperatures. Data are shown for different temperature versus field protocols (see text for details). For clarity curves are shifted by an offset. \mbox{(a3)--(b3)}~Susceptibility as a function of temperature at the applied magnetic field, for which the temperature range of the skyrmion lattice state is largest. For clarity transition fields and transition temperatures are marked for $\bm{H}\parallel\langle100\rangle$ after zero-field cooling only.}
\label{figure03}
\end{figure}

An important point of references for our observations of the magnetic structure in {\mfs} are data recorded in {\mcs}. Previous magnetization, susceptibility, resistivity and Hall effect measurements suggested that {\mfs} and {\mcs} are essentially identical when taking into account the different valence count. Namely, a cobalt content of $x_{\mathrm{Co}}$ roughly corresponds to an iron content of $x_{\mathrm{Fe}} = 2x_{\mathrm{Co}}$~\cite{2010:Bauer:PhysRevB, 2010:Teyssier:PhysRevB}. Thus, in comparison to Fe-doping the amount of defects and disorder may be expected to be a factor of two smaller under Co-doping.

The factor of two in concentration is reflected in the evolution of the anisotropy and the hysteresis of the susceptibility as illustrated in the second row of Fig.~\ref{figure03}. The general appearance of the data compare well with {\mfs}, featuring the same discrepancy of $\mathrm{d}M/\mathrm{d}H$ and $\mathrm{Re}\,\chi_{\mathrm{ac}}$ at $H_{c1}$. In particular, the helix reorientation after zero-field cooling [curve labeled (1)] is already essentially isotropic for $x_{\mathrm{Co}} = 0.02$, corresponding to the situation in Mn$_{1-x}$Fe$_{x}$Si with $x_{\mathrm{Fe}} \geq 0.04$. Likewise, after a magnetic field has been applied, marked (2) and (3), no signatures suggesting the recovery of a multi-domain character of the helical order are observed for all field directions. This implies that pinning due to defects and disorder stabilize a single-domain state under decreasing field. However, as the number of defects and disorder are smaller under Co-doping, it suggests a moderate reduction of the strength of the magneto-crystalline anisotropy potential under increasing doping in parallel to the effects of defects and disorder.

For the temperature dependence of the susceptibility, shown in Figs.~\ref{figure03}(a3) and \ref{figure03}(b3), the minimum associated with the skyrmion lattice state is also already rather isotropic in Mn$_{1-x}$Co$_{x}$Si for $x_{\mathrm{Co}} = 0.02$. However, in contrast to all other samples studied, the susceptibility at low temperatures for $x_{\mathrm{Co}} = 0.04$ is distinctly lower after field cooling down (labeled 'fcd', open symbols) as compared to data after zero-field cooling (zfc, solid symbols). This behavior is characteristic of metastable supercooling of the skyrmion lattice to low temperatures as reported, for instance, in Fe$_{1-x}$Co$_{x}$Si~\cite{2010:Munzer:PhysRevB, 2016:Bauer:PhysRevB}, Co$_{x}$Zn$_{y}$Mn$_{z}$~\cite{2016:Karube:NatMater}, MnSi under moderate hydrostatic pressure~\cite{2013:Ritz:PhysRevB}, or MnSi under violent quenching~\cite{2016:Oike:NaturePhys}. The possibility of supercooling of the skyrmion lattice is indicated by a light red shading in the phase diagram. The effect is most prominent for field along $\langle100\rangle$ but present for all major crystallographic axes. 


\section{Small-angle neutron scattering}
\label{ResultsSANS}

The main focus of our SANS measurements concerned the evolution of the spontaneous magnetic order in {\mfs} and {\mcs} and the underlying MCAs taking into account the presence of disorder. In Sec.~\ref{mfs-SANS} the SANS data and the evolution of the scattering patterns in {\mfs} are presented. These data are complemented by data recorded in {\mcs} in Sec.~\ref{mcs-SANS} which shed additional light on the role of disorder. A second aspect of the effects of disorder beyond broadening concerns the helix reorientation, of which the corresponding SANS data are presented in Sec.~\ref{Hc1-SANS}. Finally, selected data on the skyrmion lattice in {\mfs} are presented in Sec.~\ref{skx-SANS}, providing a consistency check on the formation of long-range order and MCAs as inferred from the spontaneous order. 

\subsection{Spontaneous magnetic order in Mn$_{1-x}$Fe$_{x}$Si}
\label{mfs-SANS}

\begin{figure*}
\includegraphics[width=1.0\linewidth]{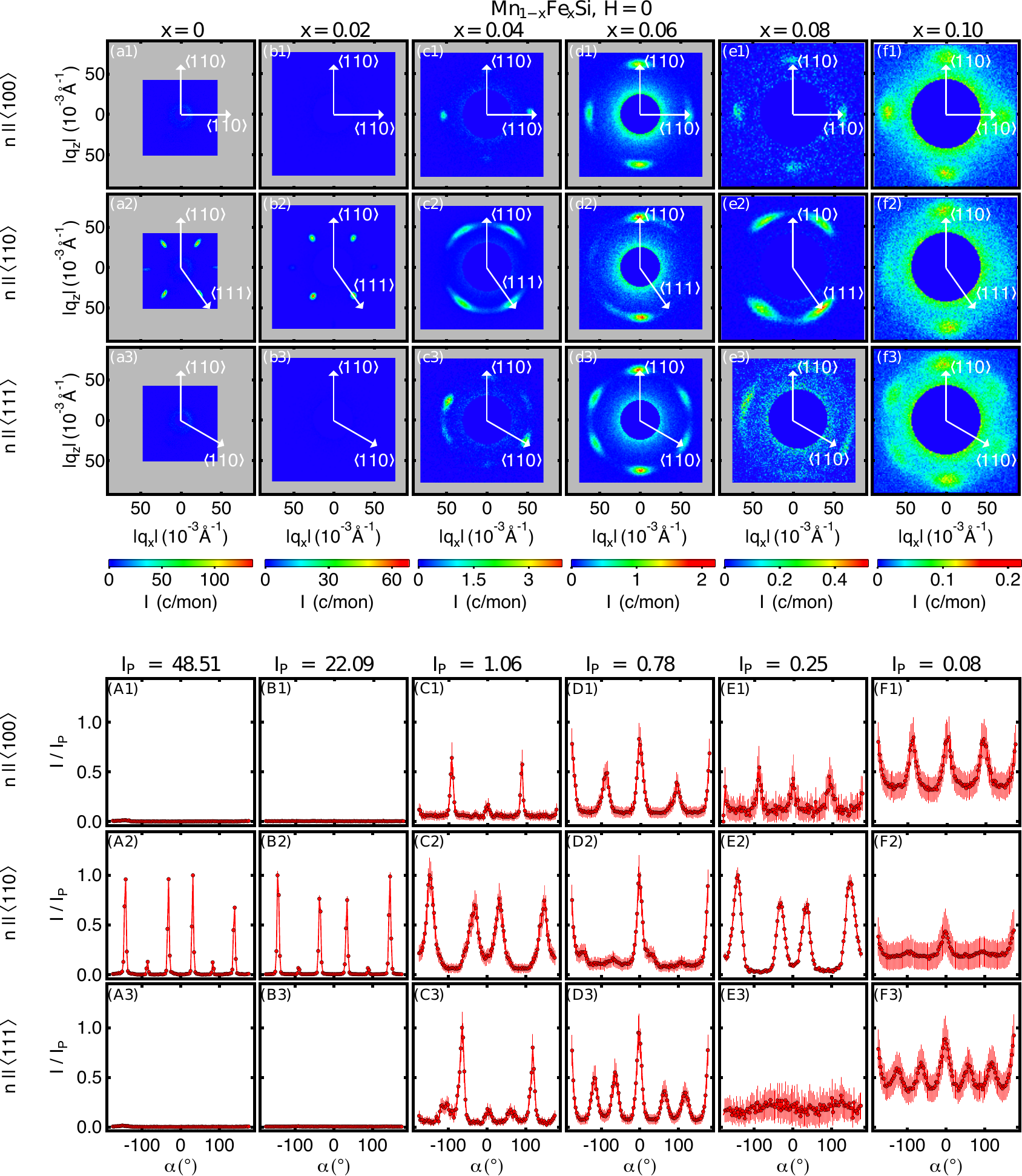}
\caption{Small-angle neutron scattering on the helical state of MnSi and Mn$_{1-x}$Fe$_{x}$Si up to $x_{\mathrm{Fe}} = 0.10$. 
In the intensity patterns sections of reciprocal space that were not covered by the neutron scattering set-up are shown in grey shading. Data for $x_{\mathrm{Fe}}\leq0.08$ were recorded in zero field at $T = 3.5$~K after zero-field cooling; data for $x_{\mathrm{Fe}} = 0.10$ were recorded at 0.5\,K. \mbox{(a1)--(a3)}~Typical sums over rocking scans for MnSi. In the first, second, and third row the incoming neutron beam, $\bm{n}$, was parallel to $\langle100\rangle$, $\langle110\rangle$, and $\langle111\rangle$, respectively. {(b1)--(f3)}~Sums over rocking scans for Mn$_{1-x}$Fe$_{x}$Si with increasing iron content from left to right. {(A1)--(F3)}~Intensity as a function of the azimuthal angle, $\alpha$, corresponding to the panels in the top three rows. The vertical $\langle110\rangle$ direction defines $\alpha = 0$. The intensity is normalized to the peak value, $I_{P}$, of each concentration.}
\label{figure04}
\end{figure*}

\begin{figure}
\includegraphics[width=1.0\linewidth]{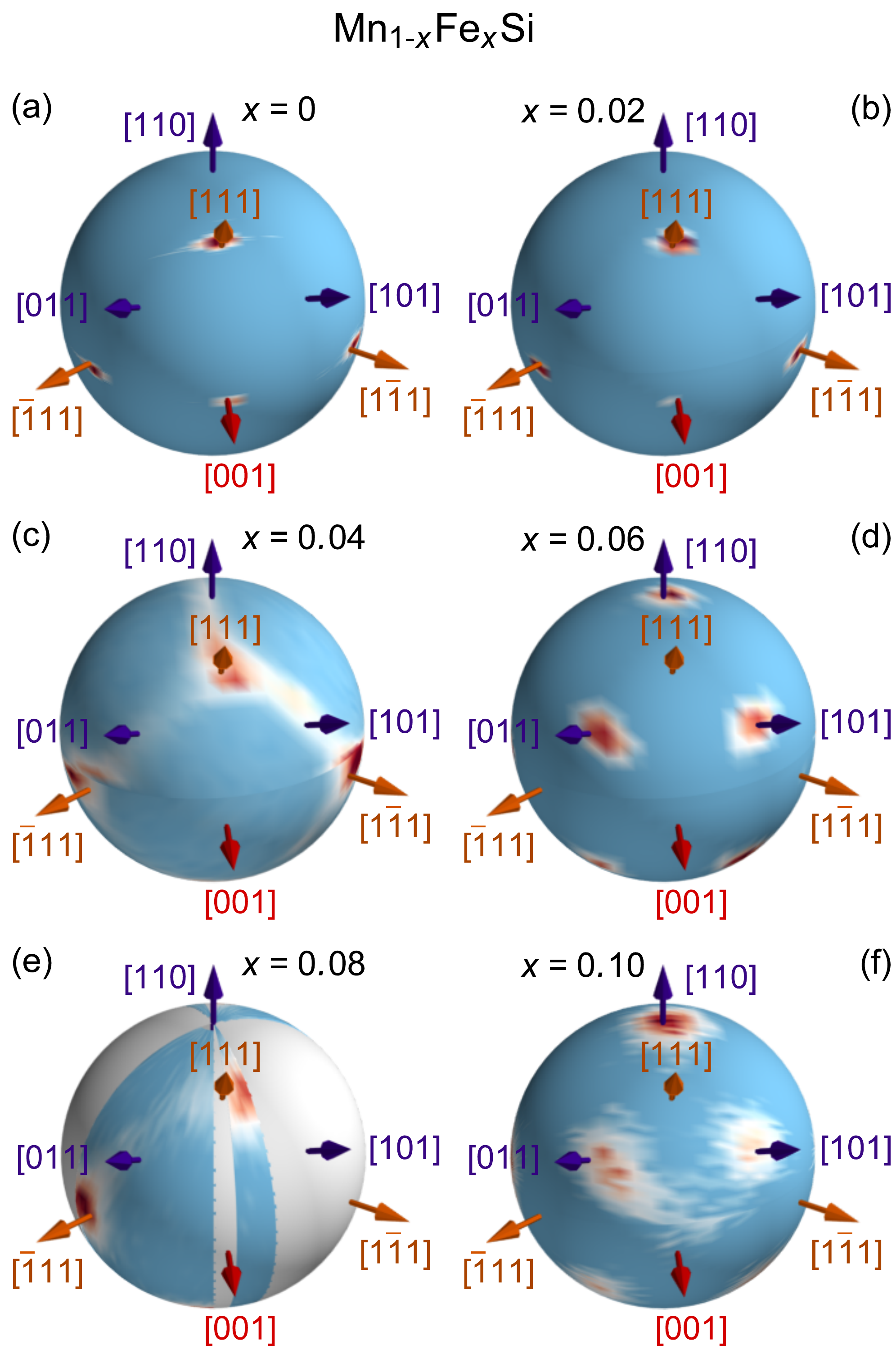}
\caption{Depiction of the experimentally observed neutron scattering intensity in zero magnetic field on spheres in reciprocal space for MnSi and Mn$_{1-x}$Fe$_{x}$Si up to $x_{\mathrm{Fe}} = 0.10$. Each sphere is constructed from the small-angle scattering data recorded for 40 different orientations of the incident neutron beam. The radius of each sphere corresponds to the modulus of the helical wave vector observed at the given composition. Angular sections for $x_{\mathrm{Fe}}=0.08$ in which no data was recorded are shown in light grey shading.}
\label{figure05}
\end{figure}

Detailed microscopic information on the spontaneous magnetic order in Mn$_{1-x}$Fe$_{x}$Si, i.e., after zero-field cooling at $H = 0$, was determined by means of small-angle neutron scattering for MnSi and Mn$_{1-x}$Fe$_{x}$Si up to $x_{\mathrm{Fe}} = 0.10$. For each composition data was recorded for 40 different orientations of the incoming neutron beam covering a sample rotation of $180^{\circ}$. Shown in Fig.~\ref{figure04} are typical intensity patterns and the corresponding azimuthal intensity distributions as recorded at $T = 3.5$~K in zero magnetic field after zero-field cooling and an incident neutron beam parallel to the three major crystallographic axis, namely $\langle100\rangle$, $\langle110\rangle$, and $\langle111\rangle$ (data for $x=0.1$ were recorded at 0.5\,K). Since the scattering patterns include broad distributions we illustrated in Fig.~\ref{figure05} this distribution of scattering weight on the surface of a sphere in momentum space with radius $Q_{\mathrm{h}}$. As a function of temperature the distribution is qualitatively unchanged. We focus therefore on the behaviour in the low-temperature limit only.

In MnSi sharp intensity maxima are observed at a wave vector $Q_{\mathrm{h}} = 0.036~\textrm{\AA}^{-1}$ along the crystalline $\langle111\rangle$ directions as shown in Figs.~\ref{figure04}(a1) to \ref{figure04}(a3), \ref{figure04}(A1) to \ref{figure04}(A3), and \ref{figure05}(a). This observation is in excellent agreement with the literature and confirms the well-established $\langle111\rangle$ easy axes for the helical modulation direction in MnSi~\cite{1976:Ishikawa:SolidStateCommun, 2006:Grigoriev:PhysRevB2, 2009:Muhlbauer:Science, 2011:Adams:PhysRevLett, 2013:Janoschek:PhysRevB, 2018:Bannenberg:SANS}. The radial width $\Delta Q = 0.004~\textrm{\AA}^{-1}$, the small azimuthal width, $\Delta\alpha = 2.74^{\circ}$, and the rocking width $\Delta\omega = 1.38^{\circ}$ are resolution-limited and characteristic of a small magnetic mosaicity of the helical order. For an incident neutron beam along $\langle100\rangle$ and $\langle111\rangle$ no intensity maxima may be seen due to the absence of $\langle111\rangle$ axes in these scattering planes, as shown in Figs.~\ref{figure04}(a1) and \ref{figure04}(a3), respectively. The weak intensity in the horizontal $\langle100\rangle$ direction for incident neutrons along $\langle110\rangle$ seen in Fig.~\ref{figure04}(a2) arises from double scattering. 

As compared with MnSi, the scattering patterns in {\mfs} feature the following major changes as a function of increasing iron content: (i)~the total scattering intensity decreases consistent with the reduction of the ordered magnetic moment~\cite{2010:Bauer:PhysRevB}, (ii)~the modulus of the helical wave vector increases, (iii)~the intensity maxima broaden, and (iv)~the easy axes of the helical modulation as inferred from the location of the intensity maxima changes. These findings are described in detail in the following. 

For $x_{\mathrm{Fe}} = 0.02$ the intensity distribution is qualitatively similar to MnSi with a moderate increase of the helical wave vector, as well as the azimuthal and the rocking width. Typical data are shown in Figs.~\ref{figure04}(b1) to \ref{figure04}(b3) and \ref{figure04}(B1) to \ref{figure04}(B3). The corresponding intensity distribution on the surface of a sphere is shown in Fig.~\ref{figure05}(b). This is followed by more substantial changes for $x_{\mathrm{Fe}} = 0.04$, where we find intensity maxima along the $\langle111\rangle$ directions accompanied by considerable broadening corresponding to an azimuthal and rocking width of $\sim10^{\circ}$ as shown in Figs.~\ref{figure04}(c1) to \ref{figure04}(c3) and \ref{figure04}(C1) to \ref{figure04}(C3). In addition weak scattering intensity may be observed for the $\langle110\rangle$ directions. As shown in Fig.~\ref{figure05}(c), the latter relates to streaks of intensity along the $\langle110\rangle$ directions connecting the maxima along the $\langle111\rangle$ directions. 

Rather surprising seems the scattering pattern for $x_{\mathrm{Fe}} = 0.06$, which displays broad intensity maxima around the $\langle110\rangle$ directions as shown in Figs.~\ref{figure04}(d1) to \ref{figure04}(d3), \ref{figure04}(D1) to \ref{figure04}(D3), and \ref{figure05}(d). In contrast, neither intensity along the $\langle111\rangle$ nor the $\langle100\rangle$ axes is observed. Moreover, in comparison to $x_{\mathrm{Fe}} = 0.04$ the intensity distribution is slightly sharper. 

For $x_{\mathrm{Fe}} = 0.08$ the intensity distribution is again qualitatively very similar to that observed for $x_{\mathrm{Fe}} = 0.04$, i.e., broad intensity maxima along the $\langle111\rangle$ directions are connected by streaks of intensity across $\langle110\rangle$ as shown in Figs.~\ref{figure04}(e1) to \ref{figure04}(e3), \ref{figure04}(E1) to \ref{figure04}(E3), and \ref{figure05}(e). For $x_{\mathrm{Fe}} = 0.10$ the scattering pattern is again reminiscent of $x_{\mathrm{Fe}} = 0.06$, exhibiting broad intensity maxima around the $\langle110\rangle$ directions as shown in Figs.~\ref{figure04}(f1) to \ref{figure04}(f3), \ref{figure04}(F1) to \ref{figure04}(F3), and \ref{figure05}(f). For $x_{\mathrm{Fe}} \geq 0.12$ (not shown) very weak uniform scattering intensity on the surface of a sphere is observed without intensity maxima in specific directions. This is consistent with the notion that long-range magnetic order vanishes above $x_{\mathrm{Fe}} = 0.10$ as proposed in several studies~\cite{2009:Grigoriev:PhysRevB}.

Taken together, the strong variations of the distribution of scattering intensities observed in small-angle neutron scattering were unexpected prior to our study and do not appear to follow a specific evolution with increasing iron concentration. Whereas intensity maxima for $\langle 111\rangle$ and $\langle 100\rangle$ may be reconciled with the previous level of description in terms of MCAs, the observation of intensity maxima along $\langle 110\rangle$ were unexpected. As discussed below, on the level that the distribution of scattering intensity changes very sensitively with composition and permitting inaccuracies of the precise value of the composition, our results are consistent with a study reporting broad intensity maxima along $\langle 111\rangle$ and $\langle 100\rangle$ for $x_{\mathrm{Fe}} = 0.08$~\cite{2009:Grigoriev:PhysRevB}.

\begin{figure}
\includegraphics[width=1.0\linewidth]{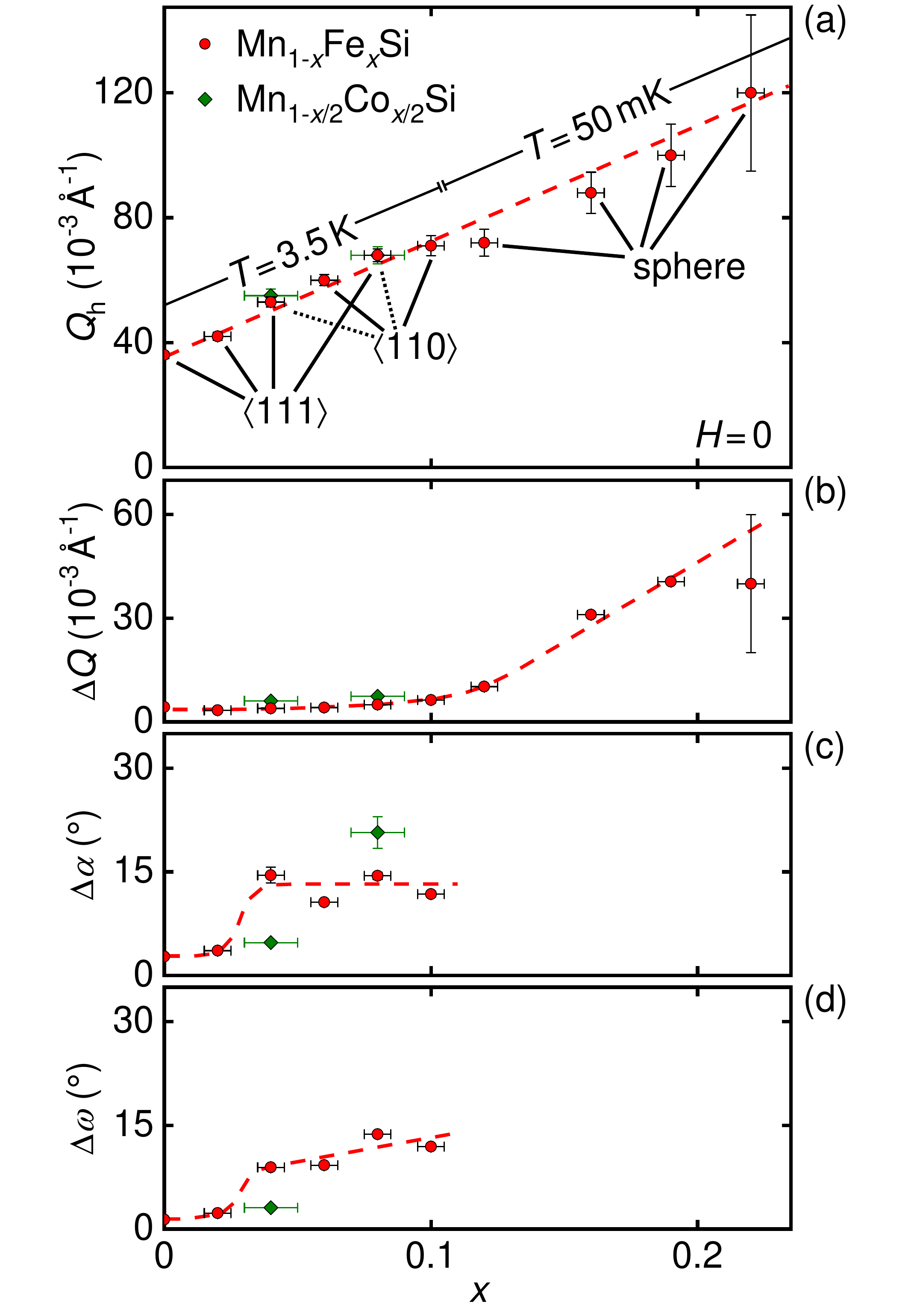}
\caption{Key parameters inferred from the SANS data recorded in {\mfs} and {\mcs} in zero field, reflecting $x_{\mathrm{Fe}}\approx 2\,x_{\mathrm{Co}}$ (see text for details) (a)~Modulus of the helical wave vector as a function of increasing iron content $x_{\mathrm{Fe}}$. For each concentration the temperature of the measurement and key characteristics of the intensity distribution are stated in the figure. Solid lines correspond to pronounced maxima, dotted lines indicate streaks of intensity. \mbox{(b)--(d)}~Radial width $\Delta Q$, azimuthal width $\Delta\alpha$, and rocking width $\Delta\omega$ as a function of increasing $x_{\mathrm{Fe}}$. Dashed lines are guides to the eye. For $x_{\mathrm{Fe}} \geq 0.12$ we observe a sphere of intensity in reciprocal space where $\Delta Q$ increases whereas $\Delta\alpha$ and $\Delta\omega$ may no longer be determined.}
\label{figure06}
\end{figure}

Key parameters characterizing the evolution of the spontaneous magnetic order in Mn$_{1-x}$Fe$_{x}$Si inferred from the SANS data are shown in Fig.~\ref{figure06}. 
This plot includes also related parameters observed in {\mcs}, where the underlying SANS data are presented in the next section. Values shown in Fig.~\ref{figure06} are plotted as a function of composition reflecting $x_{\mathrm{Fe}}\approx 2\,x_{\mathrm{Co}}$.
As a function of increasing iron content the modulus of the helical wave vector, $Q_{\mathrm{h}}$, increases linearly by a factor of 3 up to $x_{\mathrm{Fe}} = 0.22$, see Fig.~\ref{figure06}(a) in agreement with earlier reports~\cite{2009:Grigoriev:PhysRevB, 2010:Pfleiderer:JPhysCondensMatter}. As a new result we find, that $Q_{\mathrm{h}}$ under Co-doping increases in perfect agreement with $x_{\mathrm{Fe}}\approx 2\,x_{\mathrm{Co}}$. This provides strong evidence, that the DMI is due to the Berry curvature and the effects of doping well-described by rigid band splitting, i.e., the DMI is not affected by defects and disorder \cite{2013:Freimuth:PRB,2014:Freimuth:JPCM,2014:Franz:PhysRevLett}.
Moreover, despite of this decrease the exchange coupling $J$ remains larger than the DMI for all $x$. Considering, moreover, the putative evidence for a small reduction of the strength of the MCAs inferred from the properties of {\mcs} presented above, a well-defined hierarchy of energy scales exists in Mn$_{1-x}$Fe$_{x}$Si up to the highest values of $x$ studied. 

It is further instructive to keep track of the broadening of the scattering intensities of the modulus, $\Delta Q$, as well as the azimuthal and rocking widths, $\Delta\alpha$ and $\Delta\omega$, as a function of iron concentration. The radial width of the intensity maxima, $\Delta Q$, is shown in Fig.~\ref{figure06}(b). Starting from a resolution-limited value of $0.004~\textrm{\AA}^{-1}$ in MnSi, $\Delta Q$ increases very weakly for $x_{\mathrm{Fe}} \leq 0.10$ ($x_{\mathrm{Co}} \leq 0.05$). In contrast, for $x_{\mathrm{Fe}} \geq 0.12$, where uniform scattering intensity on a sphere is observed, $\Delta Q$ increases considerably as a function of increasing $x_{\mathrm{Fe}}$. Interestingly, a linear extrapolation of the increase of $\Delta Q$ with $x$ seems consistent with a suppression of long-range order for $x_{\mathrm{Fe}} \approx 0.10$ as inferred in earlier studies~\cite{2009:Grigoriev:PhysRevB}.

The azimuthal width, $\Delta\alpha$, and rocking width, $\Delta\omega$, of the intensity maxima, shown in Figs.~\ref{figure06}(c) and \ref{figure06}(d), exhibit tiny values of $\Delta\alpha = 2.7^{\circ}$ and $\Delta\omega = 1.4^{\circ}$ in MnSi. These values increase only weakly by roughly $1^{\circ}$ in both $\Delta\alpha$ and $\Delta\omega$ for $x_{\mathrm{Fe}} = 0.02$. However, for $x_{\mathrm{Fe}} \geq 0.04$ the maxima increase considerably, assuming large values of the order $10^{\circ}$. For $x_{\mathrm{Fe}} \geq 0.12$ neither $\Delta\alpha$ nor $\Delta\omega$ may be defined, as the scattering intensity is uniform on the surface of a sphere in reciprocal space. The broadening in $\Delta\alpha$ and $\Delta\omega$ up to $x_{\mathrm{Fe}} = 0.10$ shows the presence of large magnetic mosaicity. 
It is interesting to note, that $\Delta\alpha$ and $\Delta\omega$ are smaller at $x_{\mathrm{Co}} = 0.02$, which corresponds to $x_{\mathrm{Fe}} \geq 0.04$.
The dependence of $\Delta\alpha$ and $\Delta\omega$ on composition, increasing both abruptly between $x_{\mathrm{Fe}} = 0.02$ and $x_{\mathrm{Fe}} = 0.04$ and $x_{\mathrm{Co}} = 0.02$ and $x_{\mathrm{Co}} = 0.04$ suggests some influence of defect- and disorder-related pinning in the presence of a magneto-crystalline anisotropy potential as discussed below.


\subsection{Spontaneous magnetic order in Mn$_{1-x}$Co$_{x}$Si}
\label{mcs-SANS}

\begin{figure}
\includegraphics[width=1.0\linewidth]{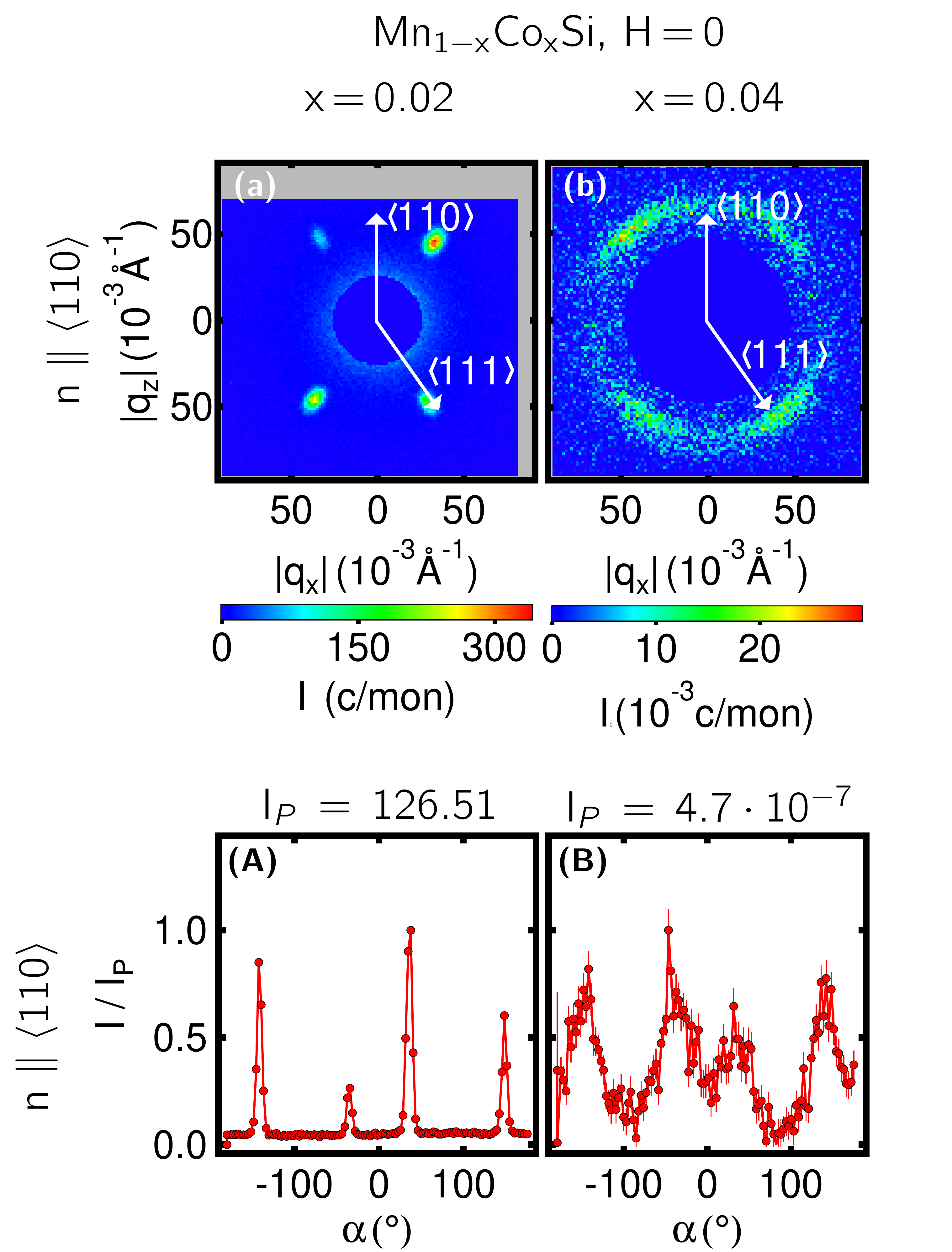}
\caption{Small-angle neutron scattering in the helical state of Mn$_{1-x}$Co$_{x}$Si with $x_{\mathrm{Co}} = 0.02$ and $x_{\mathrm{Co}} = 0.04$. Data were recorded in zero magnetic field at low temperatures ($T = 3.5$~K) after zero-field cooling. \mbox{(a),(b)}~Typical sums over rocking scans for an incoming neutron beam, $\bm{n}$, parallel to $\langle100\rangle$ and $\langle110\rangle$. \mbox{(A),(B)}~Intensity as a function of the azimuthal angle, $\alpha$, corresponding to the upper panels normalized to the peak value, $I_{P}$, of each concentration.}
\label{figure07}
\end{figure}

In the spirit of considering the bulk properties of {\mcs} as a means to gauge the effects of defects and disorder presented above, we have also recorded the microscopic magnetic properties of Mn$_{1-x}$Co$_{x}$Si by means of small-angle neutron scattering. The key parameters inferred from these data have been shown above together with the data recorded in {\mfs}. As shown in Fig.~\ref{figure07}, with increasing cobalt content the helical wave vector, as well as the azimuthal and rocking width of the intensity maxima increase. 

Consistent with previous work on the bulk properties the effect of a cobalt doping by an amount $x_{\mathrm{Co}}$ roughly corresponds to an iron doping by an amount $2x_{\mathrm{Fe}}$. For $x_{\mathrm{Co}} = 0.02$ we observe broadened intensity maxima along the $\langle111\rangle$ axes reminiscent of {\mfs} with an iron concentration of $x_{\mathrm{Fe}} = 0.04$. For {\mcs} with a cobalt concentration of $x_{\mathrm{Co}} = 0.04$ very broad maxima are observed along the $\langle111\rangle$ axes that are accompanied by intensity along the $\langle110\rangle$ directions suggesting that the $\langle111\rangle$ axes are connected by pronounced streaks of intensity in reciprocal space. This compares with {\mfs} for an iron concentration of $x_{\mathrm{Fe}} = 0.08$. The modulus of the modulation, $Q_h$, observed at $x_{\mathrm{Co}} = 0.02$ and $x_{\mathrm{Co}} = 0.04$ corresponds quantitatively to those values observed under Fe doping with $x_{\mathrm{Fe}} \approx 2\,x_{\mathrm{Co}}$.

While we have studied only two cobalt compositions, it is interesting to note that we find the same propagation directions as for the corresponding {\mfs} compositions with a factor two higher iron concentration. Yet, the broadening in {\mcs} (especially for $x_{\mathrm{Co}} = 0.02$) is smaller underscoring our conclusions, that the evolution of the spontaneous microscopic magnetic order reflects changes of the MCAs in the presence of moderate structural disorder and defects caused by the compositional doping.


\subsection{Field-induced helix reorientation in Mn$_{1-x}$Fe$_{x}$Si}
\label{Hc1-SANS}

\begin{figure}
\includegraphics[width=1.0\linewidth]{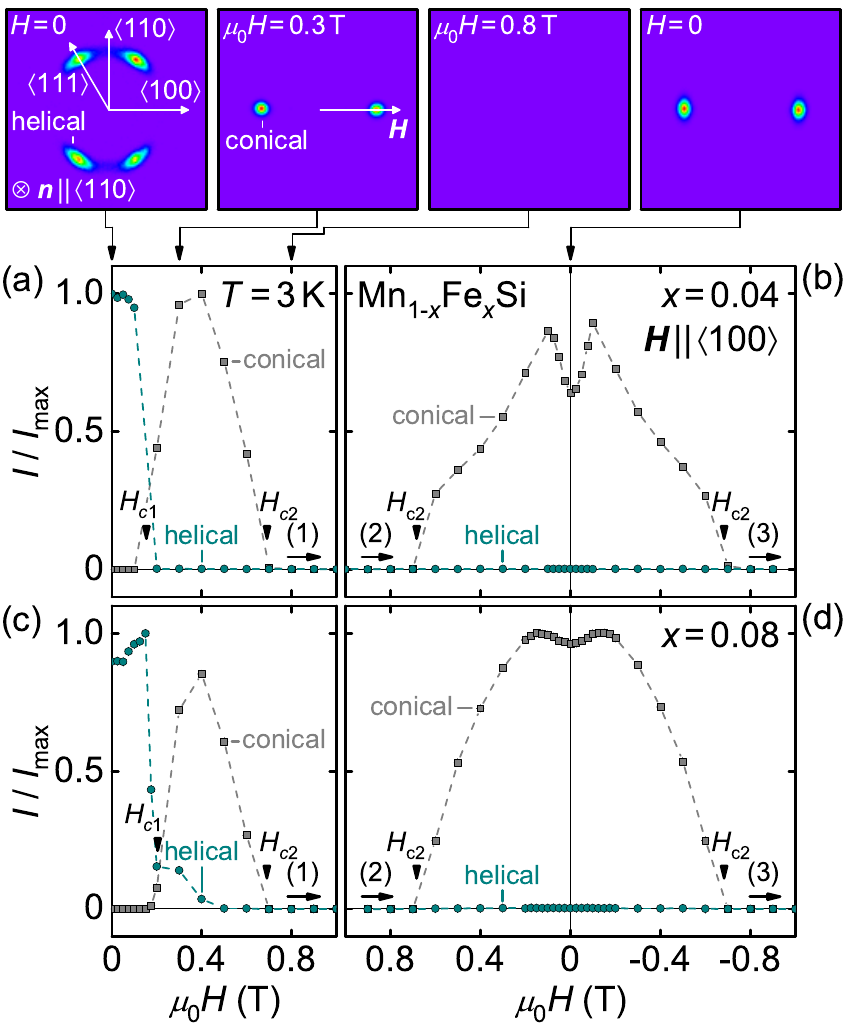}
\caption{Small-angle neutron scattering on Mn$_{1-x}$Fe$_{x}$Si as a function of magnetic field. Note the inverted field axis in panels (b) and (d) to facilitate comparison with panels (a) and (c). (a)~Intensity of the maxima corresponding to the helical and the conical state after initial zero-field cooling for $x_{\mathrm{Fe}} = 0.04$. The insets at the top show typical scattering patterns. (b)~Intensity for decreasing field values starting at $+1~\mathrm{T}$. Once the helical state has been depopulated, it is not recovered without warming to $T_{c}$. \mbox{(c),(d)}~Intensity for $x_{\mathrm{Fe}} = 0.08$ exhibiting qualitatively the same behavior.}
\label{figure08}
\end{figure}

The absence of a signature in the magnetization and susceptibility of the field-induced helix reorientation under decreasing magnetic field in {\mfs} for $x_{\mathrm{Fe}}\geq0.04$ and {\mcs} for $x_{\mathrm{Co}}\geq0.02$ when starting from field values above $H_{c1}$ suggests that pinning due to defects and disorder stabilizes a single domain state, cf.\ curves labeled (2) and (3) in Figs.~\ref{figure01} and \ref{figure03}. Field-dependent SANS measurements at low temperatures were carried out to confirm this scenario microscopically.  Data were recorded in a sequence of two field sweeps after initial zero-field cooling, first, by a sweep from zero field to $+1~\mathrm{T}$ and second, by a sweep from $+1~\mathrm{T}$ to $-1~\mathrm{T}$. To illustrate the sequence of field sweeps the corresponding data are shown from left to right in Fig.~\ref{figure08}, i.e., the field axis in Figs.~\ref{figure08}(b) and (d) is inverted to facilitate comparison with data shown in Figs.~\ref{figure08}(a) and (c). 

Typical data of the evolution of the diffraction pattern as a function of magnetic field is shown in Fig.~\ref{figure08} for $x_{\mathrm{Fe}} = 0.04$ (panels in top row). The associated integrated intensities as a function of magnetic field for $x_{\mathrm{Fe}} = 0.04$ and $x_{\mathrm{Fe}} = 0.08$ are shown in Figs.~\ref{figure08}(a) and \ref{figure08}(b) as well as Figs.~\ref{figure08}(c) and \ref{figure08}(d), respectively. Following zero-field cooling, helical order with equally populated domains is observed as presented above in Fig.~\ref{figure05}. Increasing magnetic field, as shown in Figs.~\ref{figure08}(a) and \ref{figure08}(c), stabilizes the conical state above $H_{c1}$ until the field-polarized state is reached above $H_{c2}$. 

Subsequently sweeping the magnetic field from $+1~\mathrm{T}$ to $-1~\mathrm{T}$, as shown in Figs.~\ref{figure08}(b) and \ref{figure08}(d), the conical state forms without noticeable hysteresis below $H_{c2}$. However, approaching zero field the intensity pattern remains unchanged that of the conical state until the field-polarized state is reached again below $-H_{c2}$. This is consistent with the lack of anomalies in the susceptibility at $H_{c1}$ and a single domain state. It is interesting to note a shallow minimum of the conical intensity around $H = 0$, where the intensity does not redistribute to any other location. 

We conclude that once a magnetic field exceeding $H_{c1}$ has been applied the multi-domain helicoidal state is not recovered for all field directions studied. Instead, a single-domain state is observed for $H < H_{c2}$. Since Mn$_{1-x}$Fe$_{x}$Si is subject to defects and structural disorder introduced by the substitutional doping of iron on manganese sites, an abundance of local pinning centers for helices and helical domain walls may be expected. In combination with the weak MCAs this provides a natural mechanism preventing the recovery of the multi-domain helical state once a sufficiently large magnetic field has been applied. In fact, the same defect-related pinning may also account for the smearing of the intensity maxima in azimuthal and radial direction observed at various compositions after zero-field cooling. 

It may be helpful to note that the magnetic modulation directions in a small sample volume might still rearrange themselves slightly near zero field as indicated by the tiny decrease of the conical intensity, presumably on length scales of at most a few wavelengths. Unfortunately, our small-angle neutron scattering set-up is not sensitive to such short-range effects. However, related magnetic textures were recently observed using magnetic force microscopy in Fe$_{1-x}$Co$_{x}$Si~\cite{2013:Milde:Science,2019:Milde:PhysRevB}. As both Fe$_{1-x}$Co$_{x}$Si and Mn$_{1-x}$Fe$_{x}$Si are subject to similar amounts of structural disorder, and as both compounds display similar hysteresis effects in the susceptibility~\cite{2016:Bauer:PhysRevB}, it seems plausible that the same mechanisms are active in both materials. 

It is finally also important to note that the absolute value of $H_{c1}$ in Mn$_{1-x}$Fe$_{x}$Si after zero-field cooling appears to be quite large even though all other observations consistently imply very weak crystalline anisotropies. Taking into account the strong decrease of $H_{c1}$ with increasing temperature suggestive of thermally driven unpinning of domains, the lack of orientation dependence, and the shape of the anomaly in the susceptibility at $H_{c1}$, we conclude that the field-induced helix reorientation must indeed be governed by local pinning due to defects and disorder instead of the MCAs. As a result, it is not valid to extract quantitative information on the strength of the MCAs from the value of $H_{c1}$, as previously reported in Ref.~\onlinecite{2009:Grigoriev:PhysRevB}. 


\subsection{Skyrmion lattice in Mn$_{1-x}$Fe$_{x}$Si}
\label{skx-SANS}

\begin{figure}
\includegraphics[width=1.0\linewidth]{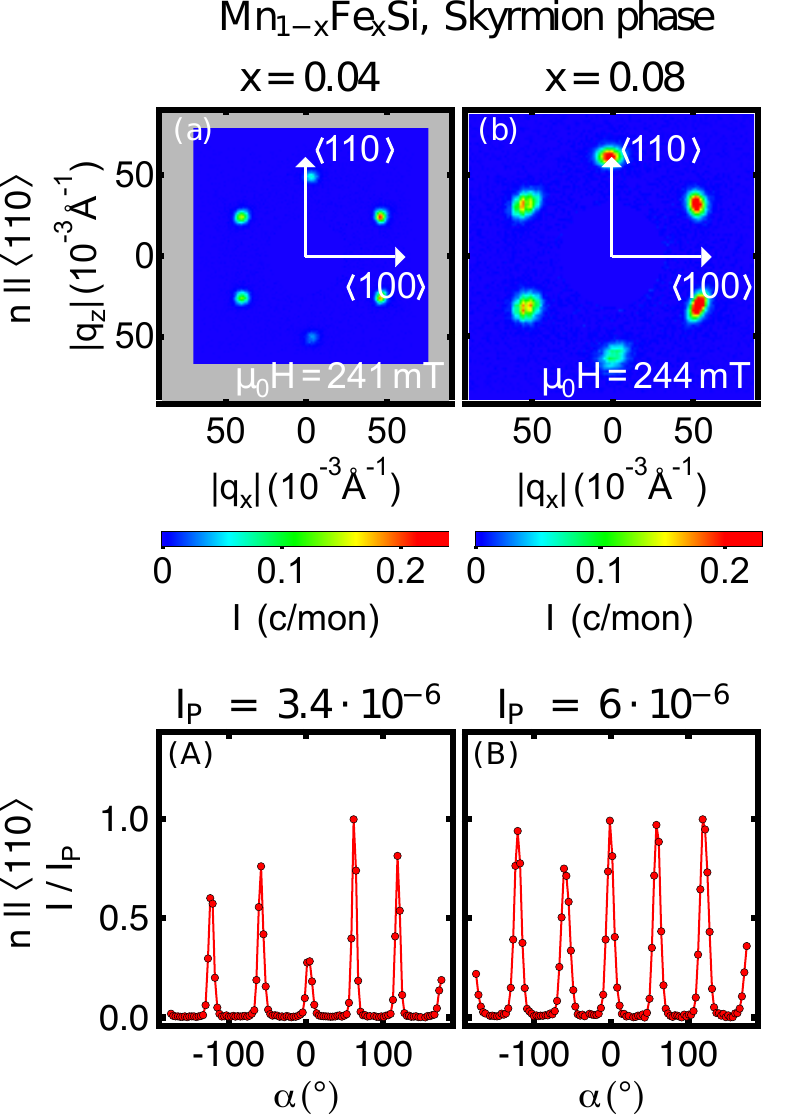}
\caption{Small-angle neutron scattering on the skyrmion lattice state of Mn$_{1-x}$Fe$_{x}$Si. (a)~Typical sixfold scattering pattern for $x_{\mathrm{Fe}} = 0.04$ for temperature and field values within the skyrmion lattice state ($T = 14.5$~K). The magnetic field was applied along the neutron beam and hence along a $\langle110\rangle$ axis. (b)~Typical sixfold scattering pattern for $x_{\mathrm{Fe}} = 0.08$ ($T = 7.5$~K). \mbox{(A),(B)}~Intensity as a function of the azimuthal angle, $\alpha$, corresponding to the upper panels normalized to the peak value, $I_{P}$, of each concentration.}
\label{figure09}
\end{figure}

High-precision SANS measurements in pure MnSi establish that the precise orientation of the skyrmion lattice with respect to the applied field and crystal lattice originates in MCAs~\cite{2018:Adams:PRL}. Contributions that are fourth order in SOC control tiny tilts of the skyrmion lattice plane against the field direction, whereas terms that are sixth order in SOC define the orientation within the skyrmion lattice plane. In order to confirm consistency of our results with the MCAs inferred at zero magnetic field, we have also tracked the evolution of the skyrmion lattice order in the doped samples. 

In the phase pocket identified by means of our susceptibility and specific heat measurements, we observe the typical sixfold scattering pattern for the magnetic field applied parallel to the neutron beam, see Fig.~\ref{figure09}. The modulus of the wave vector in the skyrmion lattice phase corresponds to that of the helical state. Data shown here do not display equal intensity for all six spots due to limitations in the rocking scans, representing a technical constraint frequently encountered. 

For $x_{\mathrm{Fe}} = 0.04$ the radial, azimuthal, and rocking width of the intensity maxima are resolution-limited, while the maxima are broadened for $x = 0.08$. In agreement with MnSi for field along $\langle110\rangle$, one of the propagation vectors of the skyrmion lattice is aligned along a $\langle110\rangle$ direction within the skyrmion lattice plane~\cite{2009:Muhlbauer:Science, 2011:Adams:PhysRevLett}. Taken together, this is consistent with the combined effect of the MCAs and disorder-induced pinning considered to account for the spontaneous magnetic order presented below. All of these findings are also consistent with early neutron scattering data on {\mfs} for $x_{\mathrm{Fe}} = 0.08$ and {\mcs} for $x_{\mathrm{Co}} = 0.04$ reported in Ref.~\onlinecite{2010:Pfleiderer:JPhysCondensMatter}.


\section{Discussion}
\label{discussion}

The discussion of our experimental results is organized in two parts. We begin with detailed considerations of the magneto-crystalline anisotropy potential in Sec.~\ref{disc-MAC}, taking into account both fourth-order and sixth-order terms in SOC. We then turn to a discussion of the different areas addressed in the introduction and the potential interest of our findings for these topics in Sec.~\ref{disc-broad}.


\subsection{Magneto-crystalline anisotropy potential}
\label{disc-MAC}

\begin{figure}
\includegraphics[width=1.0\linewidth]{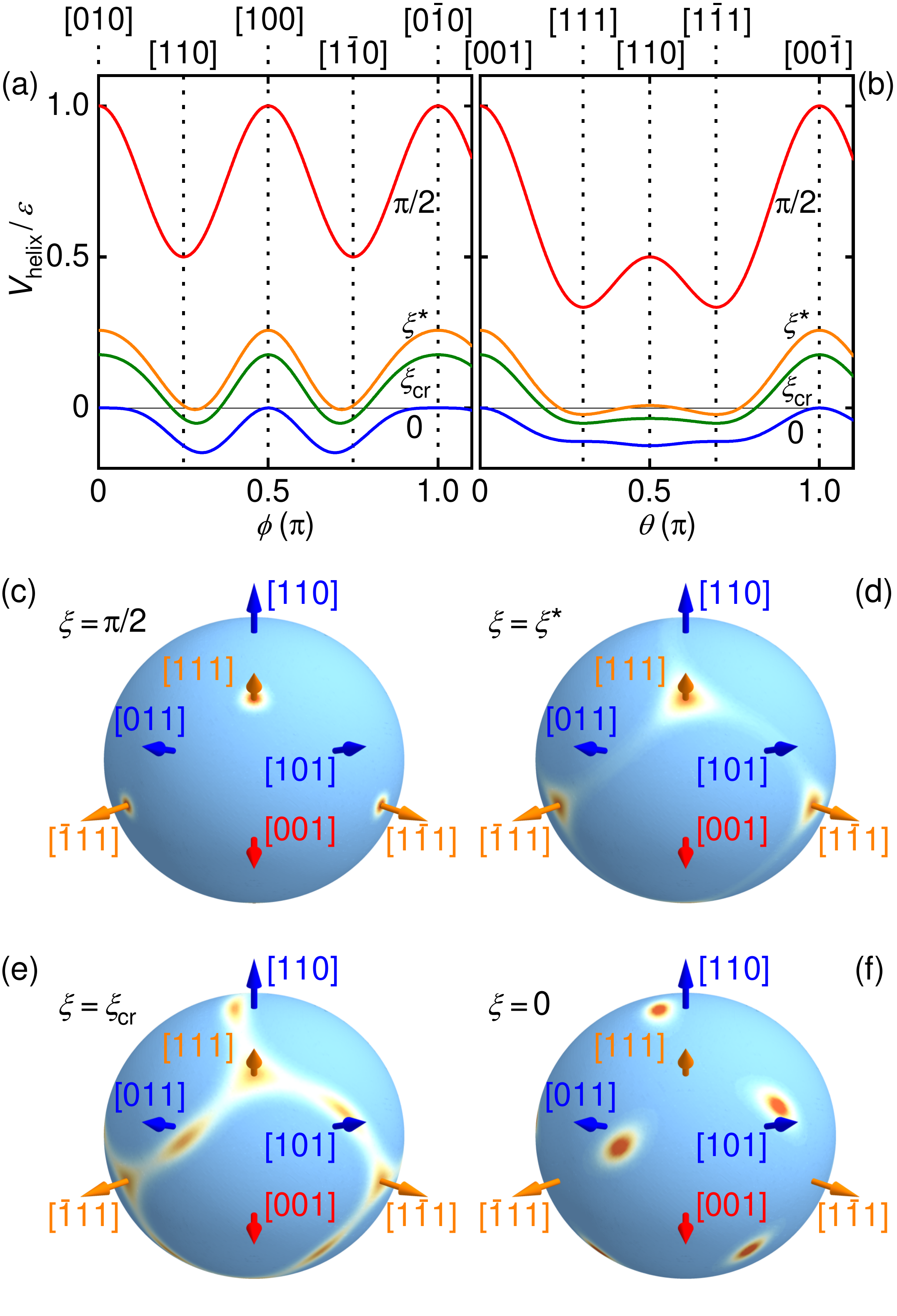}
\caption{Crystalline potential $\mathcal{V}_{\mathrm{helix}}$ for different strengths of the MCAs. The parameter $\xi$ is defined by $\varepsilon^{(1)}_{T} = \varepsilon\sin\xi$ and $\varepsilon^{(2)}_{T} = -\varepsilon\cos\xi$ with the energy density $\varepsilon > 0$. (a)~Potential for the pitch orientations $\hat{Q} = (\sin\phi,\cos\phi,0)$ as a function of the angle $\phi$ for characteristic values of $\xi$. (b)~Potential for $\hat{Q} = (\frac{\sin\theta}{\sqrt{2}},\frac{\sin\theta}{\sqrt{2}},\cos\theta)$ as a function of $\theta$. \mbox{(c)--(f)}~Boltzmann factor at finite temperature on the unit sphere for the values of $\xi$ shown in panels (a) and (b).}
\label{figure10}
\end{figure}

Our experimental results comprise two facets. On the one hand, the bulk properties display an extremely well-behaved gradual evolution of the helical transition temperature and the magnetic phase diagram as a function composition. This is characteristic of highly homogeneous samples with well-defined thermodynamic phase boundaries in the presence of weak disorder. On the other hand, the spontaneous magnetic order displays strong changes of the intensity patterns, featuring also evidence of weak disorder effects. However, considering carefully all conceivable aspects of the sample preparation as described in detail in Sec.~\ref{samples}, we could not find any hints suggesting experimental limitations such as growth-related strain or compositional gradients. Indeed, all of our findings may be fully accounted for theoretically by the usual MCAs as follows.

The well-known phenomenological model of the hierarchy of energy scales in cubic chiral magnets is based on a Ginzburg--Landau ansatz of the free energy density comprising two contributions, $f = f_{0} + f_{\mathrm{cub}}$. The first term accounts for the exchange and the Dzyaloshinsky--Moriya interactions as well as the Zeeman coupling to an externally applied magnetic field and reads in the notation of Ref.~\onlinecite{2013:Janoschek:PhysRevB}
\begin{equation}
\label{FreeEnergy}
\begin{split}
f_{0} = \frac{1}{2}\bm{\phi}(r-J\nabla^{2})\bm{\phi} &+ D\bm{\phi}(\nabla \times \bm{\phi}) + \nonumber\\*
&+ \frac{u}{4!}(\bm{\phi}^{2})^{2} - \mu_{0}\mu\bm{\phi}\bm{H}.
\end{split}
\end{equation}
Here, $\phi$ is the three-component order parameter field describing the dimensionless magnetization and $r$ tunes the distance to the phase transition, while $J$ and $u$ are the stiffness and the interaction parameter of the ferromagnetic exchange. $D$ corresponds to the Dzyaloshinsky--Moriya coupling constant that is proportional to spin--orbit coupling $\lambda_{\mathrm{SOC}}$ and $\bm{H}$ is the magnetic field. The second term of the free energy density $f_{\mathrm{cub}}$ accounts for the crystalline anisotropies and breaks rotation symmetry already in zero field. Typically, it either favors easy $\langle100\rangle$ or $\langle111\rangle$ axes for the propagation vector of the helix~\cite{1980:Bak:JPhysCSolidState}.

As discussed in detail in Ref.~\onlinecite{2017:Bauer:PRB}, accurate measurements of the helix reorientation in MnSi establish the effective magneto-crystalline anisotropy potential $\mathcal{V}_{\mathrm{helix}}$ for the pitch orientation, i.e., the unit vector $\hat{Q}=\vec{Q}/\vert \vec{Q}\vert$, as
\begin{equation}
\begin{split}
\label{PitchPotential}
\mathcal{V}_{\mathrm{helix}}(\hat{Q}) &= \varepsilon^{(1)}_{T} (\hat{Q}_{x}^{4} + \hat{Q}_{y}^{4} + \hat{Q}_{z}^{4}) + \nonumber\\*
&+ \varepsilon^{(2)}_{T} (\hat{Q}_{x}^{2} \hat{Q}_{y}^{4} + \hat{Q}_{y}^{2} \hat{Q}_{z}^{4} + \hat{Q}_{z}^{2} \hat{Q}_{x}^{4}) + ...~.
\end{split}
\end{equation}
Since the magnitude of the pitch $Q = D/J \propto \lambda_{\mathrm{SOC}}$ varies with spin--orbit coupling $\lambda_{\mathrm{SOC}}$, the first and second term are expected to scale as a power of $\lambda^{4}_{\mathrm{SOC}}$ and $\lambda^{6}_{\mathrm{SOC}}$, respectively. As a consequence, the first term in $\varepsilon^{(1)}_{T}$ may be expected to dominate generically, determining the orientation of the helix. In particular, the first term favors either $\langle100\rangle$ easy axes for $\varepsilon^{(1)}_{T} < 0$, as observed in Cu$_{2}$OSeO$_{3}$~\cite{2012:Adams:PhysRevLett, 2012:Seki:PhysRevB}, or $\langle111\rangle$ easy axes for $\varepsilon^{(1)}_{T} > 0$, as observed in MnSi~\cite{1976:Ishikawa:SolidStateCommun}. A transition of the easy axis as a function of temperature has been reported in FeGe, consistent with a change of sign of $\varepsilon^{(1)}_{T}$~\cite{1989:Lebech:JPCM}. We have no evidence that such changes of MCA as a function of temperature are present in {\mfs} or {\mcs}. In the spirit of a change of magneto-crystalline anisotropy potential as a function of composition, the scattering pattern observed in {\mfs} for $x_{\mathrm{Fe}} = 0.08$ has been interpreted as to provide putative evidence of a reduction of $\varepsilon^{(1)}$ under iron doping without further considerations of the next higher terms~\cite{2009:Grigoriev:PhysRevB}.

However, it is important to note that the term in $\varepsilon^{(1)}$ possesses $C_{4}$ rotation symmetry around the cubic axes that is not present in the tetrahedral point group $T$ of the crystal structure. This symmetry is explicitly broken down to $C_{2}$ symmetry by the next-leading-order term in $\varepsilon^{(2)}_{T}$. A detailed study of the helix reorientation in MnSi establishes that the second term given here is essential for a full account of the pitch reorientation transition as a function of increasing magnetic field~\cite{2017:Bauer:PRB}. In comparison, further terms that are sixth order in SOC, such as $\hat{Q}_{x}^{2}\hat{Q}_{y}^{2}\hat{Q}_{z}^{2}$ and $(\hat{Q}_{x}^{6} + \mathrm{cycl.})$, preserve the $C_{4}$ symmetry. In the following we neglect these terms for simplicity.

In case the first term in $\varepsilon^{(1)}_{T}$ becomes very small or vanishes, the second term $\varepsilon^{(2)}_{T}$ may control the helix orientation in zero magnetic field favoring either $\langle100\rangle$ easy axes for $\varepsilon^{(2)}_{T} > 0$ or, interestingly, easy axes close to $\langle110\rangle$ for $\varepsilon^{(2)}_{T} < 0$. This scenario, i.e., that $\varepsilon^{(1)}_{T} > 0$ is tuned towards zero so that the pitch orientation is qualitatively influenced by the second term $\varepsilon^{(2)}_{T}$, offers a minimal explanation for the observed intensity distributions in Mn$_{1-x}$Fe$_{x}$Si, as presented in further detail in the following.

As the focus of our study concerns the observation of intensity distributions including gradual variations, we have calculated the potential landscape as a function of orientation for selected parameters, shown in Figs.~\ref{figure10}(a) and \ref{figure10}(b), and the associated scattering patterns as depicted on the surface of a sphere, shown in Figs.~\ref{figure10}(c) to \ref{figure10}(f). The former serves to document the precise evolution of the minima and maxima of the potential, whereas the latter serves to permit direct comparison with the experimental results shown in Fig.~\ref{figure05}. For the sake of ease of presentation and ease of discussion, we have parametrized the potential $\mathcal{V}_{\mathrm{helix}}$ here in terms of $\varepsilon^{(1)}_{T} = \varepsilon\sin\xi$ and $\varepsilon^{(2)}_{T} = -\varepsilon\cos\xi$ for various values, where $\xi$ represents a tuning parameter and $\varepsilon$ represents the total strength of the potential with $\varepsilon > 0$. 

Fig.~\ref{figure10}(a) displays the potential for pitch orientations $\hat{Q} = (\sin\phi,\cos\phi,0)$ as a function of $\phi$, while Fig.~\ref{figure10}(b) displays the potential for pitch orientations $\hat{Q} = (\frac{\sin\theta}{\sqrt{2}},\frac{\sin\theta}{\sqrt{2}},\cos\theta)$ as a function of $\theta$. For this choice of parameters, the potential exhibits maxima along the cubic $\langle100\rangle$ axes and minima along the $\langle111\rangle$ directions. The latter represent global minima for $\pi/2 \geq \xi > \xi_{\mathrm{cr}} \approx 0.18$. Note that for $\xi = \pi/2$ the $\langle110\rangle$ axes represent saddle points. However, if the first term of $\mathcal{V}_{\mathrm{helix}}$ in $\varepsilon^{(1)}_{T}$ becomes sufficiently small assuming a value in the range $0 \leq \xi < \xi^{*} \approx 0.26$ a second set of minima close to $\langle110\rangle$ emerges. In fact, these minima become global as a function of $\xi$ at $\xi_{\mathrm{cr}}$ by virtue of a first-order phase transition. The positions of the new minima are not exactly at $\langle110\rangle$, but only close to these directions. In the limit $\xi = 0$, the new minima shift towards $[1\sqrt{2}0]$ and directions equivalent with respect to the tetrahedral point group of the crystal structure. Here further terms that are sixth order in SOC may become relevant for the full account of the experimental results, which is, however, beyond the scope of our study.


Further, close to the phase transition as a function of $\xi$, the potential between the $\langle111\rangle$ and $\langle110\rangle$ directions is very shallow. This finding is illustrated in Figs.~\ref{figure10}(c) through \ref{figure10}(f) where we show the Boltzmann factor $\exp[-\mathcal{V}_{\mathrm{helix}}\xi_{\mathrm{dom}}^{3} / (k_{\mathrm{B}}T)]$ on the unit sphere in order to compare the results of our calculations directly with the neutron scattering data shown in Fig.~\ref{figure05}. Here, $\xi_{\mathrm{dom}}$ represents the linear size of a helical domain, where $k_{\mathrm{B}}$ is the Boltzmann constant, and $T$ is the temperature. For the theoretical results presented here we set $k_{\mathrm{B}}T / (\xi_{\mathrm{dom}}^{3}\varepsilon) = 0.005$. For $\xi = \pi/2$ ($\varepsilon^{(2)}_{T} = 0$) the weight of the scattering intensity is concentrated at $\langle111\rangle$, cf.\ Fig.~\ref{figure10}(c). With decreasing $\xi$ streaks towards $\langle110\rangle$ appear around $\xi^{*}$, cf.\ Fig.~\ref{figure10}(d). This weight of intensity is redistributed at the phase transition, cf.\ Fig.~\ref{figure10}(e), and is concentrated at $[1\sqrt{2}0]$ and equivalent positions for $\xi = 0$ ($\varepsilon^{(1)}_{T} = 0$), cf.\ Fig.~\ref{figure10}(f).

The excellent qualitative agreement of the minimal magneto-crystalline anisotropy potential with our experimental results suggests strongly that in Mn$_{1-x}$Fe$_{x}$Si $\xi \approx \pi/2$ up to $x_{\mathrm{Fe}} \leq 0.02$ and hence $\varepsilon^{(1)}_{T} \gg \varepsilon^{(2)}_{T}$. In contrast, for $x_{\mathrm{Fe}} = 0.04$ and $x_{\mathrm{Fe}} = 0.08$ we find $\xi \approx \xi^{*}$, while $\xi < \xi_{\mathrm{cr}}$ applies for $x_{\mathrm{Fe}} = 0.06$ and $x_{\mathrm{Fe}} = 0.10$. 
The same is true for {\mcs} with $x_{\mathrm{Fe}}\approx 2\,x_{\mathrm{Co}}$. The broadening of the intensity maxima implies further a tiny value of the potential strength $\varepsilon$. As already mentioned above, we note that our experimental data do not allow to unambiguously distinguish between propagation along $\langle110\rangle$ and directions close to the latter as suggested by the potential $\mathcal{V}_{\mathrm{helix}}$. However, considering the agreement of the total number of scattering maxima observed experimentally with the magneto-crystalline anisotropy potential considered here, namely eight for the case of $\langle111\rangle$ easy axes and twelve for the case of $\langle110\rangle$ easy axes, suggests strongly that the key aspects are captured appropriately. A quantitative comparison, however, will require a neutron scattering study with high angular resolution.

The very shallow nature of the anisotropy landscape in Mn$_{1-x}$Fe$_{x}$Si, which corresponds to a weak energy scale, is also corroborated by the neutron scattering results reported in Ref.~\onlinecite{2009:Grigoriev:PhysRevB}. For {\mfs} with $x_{\mathrm{Fe}} = 0.08$ this study reported broad maxima along $\langle100\rangle$ and $\langle111\rangle$ connected by streaks of intensity. While the qualitative character of this intensity distribution is similar to those observed in our study, when permitting for inaccuracies of composition, the positions of the maxima in reciprocal space differ. Such an intensity distribution may be associated, for instance, with a small but finite value of $\varepsilon^{(1)}_{T}$ that changes its sign across the sample due to inhomogeneities of some kind.


While the magneto-crystalline anisotropy potential we consider here accounts for the intensity patterns, it cannot explain variations of $\xi$ with composition. In particular, one might expect a monotonic evolution, whereas we observe non-monotonic variations. As summarized in section \ref{ExpMeth} we have exercised great efforts to rule out extrinsic mechanisms such as compositional gradients or grow-in strain as the cause of these non-monotonic variations of $\xi$ with composition. Moreover, we have confirmed in several independent measurements the orientation of our samples. In combination with the monotonic evolution of the transition temperature, ordered magnetic moment, susceptibility, specific heat, transport properties\cite{2014:Franz:PhysRevLett}, and modulation wave length as a function of composition, all of which are characteristic of an excellent sample quality and compositional homogeneity, we are forced to consider an intrinsic mechanism that controls $\xi$. Indeed, ab initio calculations reported in Ref.\,\onlinecite{2014:Franz:PhysRevLett} suggest that the effects of doping on the Fermi surface may be approximated well by a rigid band splitting of the Fermi surface of pure MnSi that matches the ordered moment under doping. In turn, the changes of sign might be directly related to subtle intrinsic details of the Fermi surface. This identifies the need for high-resolution ab initio calculations of the magneto-crystalline anisotropies in {\mfs} and {\mcs}.


\subsection{Broader implications}
\label{disc-broad}
 
The strong changes of intensity patterns with composition we observe in {\mfs} and {\mcs} are consistent with cubic MCAs, where the character changes from contributions that are fourth-order to terms that are sixth-order in SOC. It compares at the same time with the lack of orientation dependence of the susceptibilities at the helix reorientation, the decrease of $H_{c1}$ with increasing temperature suggestive of thermally driven unpinning of domains, and the absence of the signature of the helix reorientation in the magnetization and susceptibility once the magnetic field exceeded $H_{c1}$ together with the observation of single domain formation in SANS. All of these findings suggest consistently that the behavior at $H_{c1}$ is dominated by defects and disorder. In turn, our findings question previous attempts to infer the strength of the MCAs in {\mfs} from the value of $H_{c1}$~\cite{2009:Grigoriev:PhysRevB}.  

Moreover, while the sixth-order contributions may be expected to dominate the details of the helix reorientation processes in {\mfs} and {\mcs} under magnetic field, the effects of disorder clearly prevent more detailed examination such as those reported recently for MnSi~\cite{2017:Bauer:PRB}. Instead, the in-plane orientation of the skyrmion lattice as reported here provides direct information on the sixth-order terms in SOC. However, it is important to note that high-precision measurements of the skyrmion lattice alignment require great efforts to avoid signal contamination by demagnetizing fields as observed in early work on MnSi~\cite{2011:Adams:PhysRevLett, 2018:Reimann:PhysRevB, 2018:Adams:PRL}. They are therefore beyond the scope of the work reported here.

An important question concerns the compatibility of the shift of character of the MCAs from terms that are fourth to sixth order in SOC with the microscopic origin of the hierarchy of energy scales. As summarized in Sec.\,\ref{microsc}, recent measurements of the ordinary, anomalous and topological Hall effects and ab-initio calculations in {\mfs} demonstrated that the doping and pressure dependence of the electronic structure is well accounted for by a rigid band splitting~\cite{2014:Franz:PhysRevLett}. The quantitative consistency of $\vert \vec{Q}\vert$ and the Hall effects between {\mfs} and {\mcs}, where $x_{\rm Fe}=2\,x_{\rm Co}$, underscores that the ferromagnetic exchange and the DMI originate in the electronic structure and the Berry curvature subject to rigid band splitting under doping.

Provided the MCAs are related to degenerate points of the Fermi surface as may be generic in the class of weak itinerant-electron magnets \cite{sigfusson:82,kondorsky:74,2008:Buenmann:PhysRevLett}, it is conceivable that such degeneracies are especially sensitive to defect- and disorder-related scattering and tiny modifications. In particular, it seems plausible that defects generate a smearing of fine details of the electronic structure especially at points featuring degeneracies such that the character of the MCAs is affected without changing the overall strength significantly. Unfortunately, the theoretical and experimental work required to verify such a mechanism are beyond the scope of the study presented here.

Regarding the QPT in MnSi and FeGe under hydrostatic pressures the results of our study offer an unexpected explanation for the observation of the partial magnetic order. While the broad intensity maxima for the $\langle110\rangle$ axes appeared to be a mystery for a long time, we find here that they may be explained by a shift of the character of the MCAs where terms that are sixth order in SOC become important. This was so far believed to be unlikely.
However, it is conceivable that a related modification of fourth- and sixth-order SOC terms is also present under pressure. In fact, such a scenario would also be consistent with a great sensitivity to non-hydrostatic pressure conditions reported in the literature~\cite{2004:Pfleiderer:Nature, 2005:Fak:JPhysCondMatter, 2007:Pfleiderer:PhysRevLett, 2019:Bannenberg:PhysRevB}, as well as the effects of uniaxial pressure\cite{2015:Chacon:PhysRevLett,2015:Nii:NatCommun}.

It is further important to note that a simple account for the broad intensity maxima in the $\langle 110\rangle$ directions of the partially ordered state such as the changes of the MCAs observed here, does not rule out dominant topological spin textures as a key characteristic of the non-Fermi liquid behavior \cite{2013:Ritz:Nature}. Indeed, stabilization of skyrmions by means of entropic effects associated with an abundance of fluctuations does not require MCAs, which represents an independent stabilization mechanism \cite{2018:Chacon:NatPhys,2018:Halder:PRB,1989:Bogdanov,1994:Bogdanov}.

Our observations on the MCAs in {\mfs} and {\mcs} offer, moreover, a simple explanation for observations reported in Fe$_{1-x}$Co$_{x}$Si. SANS measurements in {\fcs} with $x_{\mathrm{Co}} = 0.20$ revealed here an intensity distribution after zero-field cooling with intensity everywhere on the surface of a small sphere in reciprocal space featuring broad maxima along $\langle110\rangle$, i.e., an intensity distribution akin to the partial order in MnSi under pressure~\cite{2010:Munzer:PhysRevB}. Again small modifications of the electronic structure near degeneracies of the Fermi surface may change the character of the MCAs. 

\begin{figure}
\includegraphics[width=1.0\linewidth]{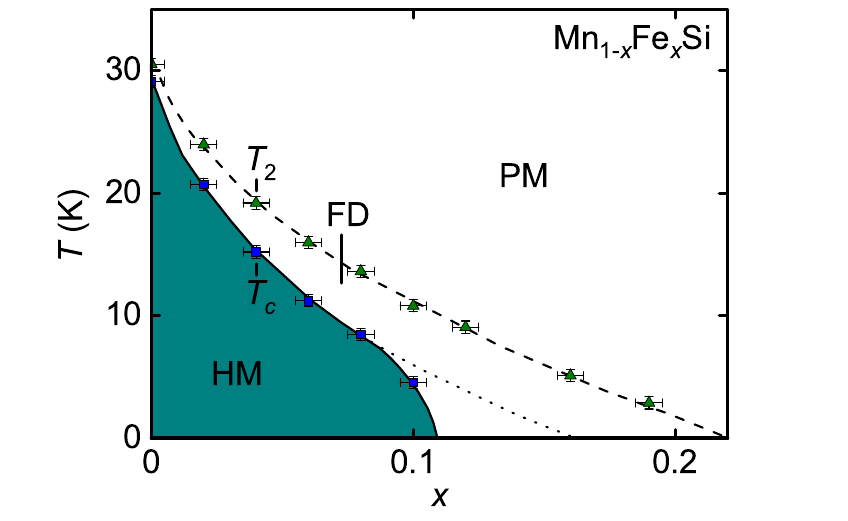}
\caption{Compositional phase diagram of Mn$_{1-x}$Fe$_{x}$Si. With increasing iron content, $x_{\mathrm{Fe}}$, long-range helimagnetic order is suppressed while the magnetic phase diagram remains qualitatively very similar to pure MnSi. The fluctuation-induced first-order transition at $T_{c}$ marks the onset of helimagnetic order~(HM). The Vollhardt invariance at $T_{2}$ tracks the crossover between the fluctuation-disordered~(FD) and the paramagnetic regime~(PM). In comparison to  previous phase diagrams reported in the literature (cf. Ref.\,\onlinecite{2010:Bauer:PhysRevB}) the behaviour near the critical composition has been amended.}
\label{figure11}
\end{figure}

On a final note, it is also instructive to revisit the temperature versus composition phase diagram of {\mfs} shown in Fig.~\ref{figure11}. An important amendment as compared to the literature \cite{2010:Bauer:PhysRevB}, concerns the modification of the phase boundary in the vicinity of the critical concentration between $x = 0.10$ and $x = 0.12$, where static magnetic order inferred from the results presented in this study is suppressed. This highlights also the putative existence of fluctuating textures or quantum Brazovskii scenario in the regime below the dotted line asking for further studies beyond the scope of the work reported here.

Starting point of the Brazovskii scenario in pure MnSi has been the observation of small MCAs \cite{2013:Bauer:PhysRevLett, 2016:Stishov:PRB}. While recent studies claim to observe differences to the Brazovski scenario  \cite{Pappas:PRL2017}, they do not question the strength of the MCAs and do not offer an alternative interpretation. In turn, the change of character of the MCAs without pronounced change of overall strength under increasing concentration we report here, is perfectly consistent with the presence of the fluctuation-disordered~(FD) regime and the associated Brazovskii scenario all the way up to high concentrations. However, under increasing doping the properties at high concentrations may eventually be dominated by disorder. This has been considered in a number of studies on polycrystalline samples, where short-range order and glassy behavior have been reported~\cite{2010:Teyssier:PhysRevB, Demishev:2016}.


\section{Conclusions}
\label{Conclusion}

Combining comprehensive small-angle neutron scattering with measurements of the magnetization, ac susceptibility, and specific heat, we have tracked the evolution of the magnetic properties of {\mfs} over a wide range of compositions. The experimental observations and conjectures drawn from these data are consistent with measurements in {\mcs} as recorded for selected compositions. 

In accord with the literature, the thermodynamic phase diagram evolves monotonically with increasing $x_{\mathrm{Fe}}$ ($x_{\mathrm{Co}}$), where long-range helimagnetic order is suppressed for a concentration $x_{\mathrm{Fe}} > 0.10$ ($x_{\mathrm{Co}} > 0.05$)~\cite{2009:Grigoriev:PhysRevB, 2010:Bauer:PhysRevB, 2010:Pfleiderer:JPhysCondensMatter}. As a new result we find, that the helix reorientation at $H_{c1}$ for $x_{\mathrm{Fe}} \geq 0.04$ ($x_{\mathrm{Co}} \geq 0.02$) becomes isotropic, and exhibits a temperature dependence characteristic of thermally driven unpinning, while the signatures of the helix reorientation, once magnetized, are suppressed in the magnetization and susceptibility consistent with single-domain formation observed in SANS. This clearly establishes that the helix reorientation under doping reflects the effects of defects and disorder and may no longer be used to infer the strength of the cubic MCAs. Yet, the observation of specific heat anomalies at the boundary of the skyrmion lattice phase shows that the effects of defects and disorder are still rather weak and similar to the MCAs, i.e., the weakest scale.

Carefully mapping the spontaneous SANS intensity distribution at zero field revealed, moreover, considerable changes of the spontaneous (zero-field) intensity distributions in the helimagnetic state as a function of increasing $x$. In particular, for {\mfs} with $x_{\mathrm{Fe}} = 0.06$ and $x_{\mathrm{Fe}} = 0.10$ broad intensity maxima around the $\langle110\rangle$ directions are observed, which cannot be explained with MCA terms that are fourth order in SOC, considered in the literature so far. Instead, all of our findings may be explained when taking into account contributions of the MCAs that are sixth order in SOC. 

Recognizing the importance of MCAs in {\mfs} and {\mcs} that are sixth order in SOC as compared to terms that are fourth order in SOC is of interest to a wide range of topics. This concerns the generic magnetic phase diagram, the paramagnetic-to-helimagnetic transition, and the morphology of topological spin textures. In particular, it offers a simple explanation for the observation of broad intensity maxima for the $\langle110\rangle$ directions in the partially ordered state of MnSi at high pressures. Thus, our study connects the complex evolution of the magnetic order in MnSi under pressure with those under Fe and Co doping, shedding new light on the possible nature of partial magnetic order and the QPT in these and related systems. Keeping track of the magnetic anisotropies vis a vis the comparatively  small amount of defects and disorder, our results underscore the putative importance of the Brazovskii scenario for the QPTs in {\mfs}, {\mcs} and related compounds.


\acknowledgments

We wish to thank P.\ B\"{o}ni, C.\ Franz, M.\ Halder, S.\ Mayr, F.\ Rucker, and C.\ Schnarr for fruitful discussions and assistance with the experiments. Financial support through DFG TRR80 (projects E1 and F7), DFG FOR960 (Quantum Phase Transitions), ERC Advanced Grant 291079 (TOPFIT), ERC Advanced Grant 788031 (ExQuiSid), and DFG SPP2137 (Skyrmionics) is gratefully acknowledged. J.K., T.A., A.B., F.H., and A.C.\ acknowledge financial support through the TUM graduate school.

\end{document}